\documentclass[a4paper,aps,11pt,nofootinbib,pdftex]{revtex4}
\usepackage{amsmath}
\usepackage{amssymb}
\usepackage{amsfonts}
\usepackage{color}
\usepackage{float}
\usepackage{comment}
\usepackage{ulem}
\usepackage{ascmac}
\usepackage{graphicx}
\begin{document}

%%%%%%%%%%%%%%%%%%%%%%%%%%%%%%%%%%%%%%%%
\title{%Numerical Research of
Perfect Charge Screening of Extended Sources \\
in an Abelian-Higgs Model
\vspace{1cm}
}
%\today

\hfill{OCU-PHYS 490}

\hfill{AP-GR 150}

%\pacs{04.50.Gh}

\author{Hideki~Ishihara}\email{ishihara@sci.osaka-cu.ac.jp}
\author{Tatsuya Ogawa}
\email{taogawa@sci.osaka-cu.ac.jp}
\affiliation{
 Department of Mathematics and Physics,
 Graduate School of Science, Osaka City University,
 Osaka 558-8585, Japan}

%%%%%%%%%%%%%%%%%%%%%%%%%%%%%%%%%%%%%%%%
\begin{abstract}
\vspace{1cm}
We investigate a classical system that consists of a U(1) gauge field and a complex
Higgs scalar field with a potential that breaks the symmetry spontaneously.
We obtain numerical solutions of the system in the presence of a smoothly
extended external source with a finite size.
In the case of the source is spread wider than the mass scale of the gauge field, perfect screening
of the external source occurs, namely, charge density of the source is canceled out everywhere
by induced counter charge density cloud of the scalar and vector fields.
Energy density induced by the cloud is also obtained.

\end{abstract}
%%%%%%%%%%%%%%%%%%%%%%%%%%%%%%%%%%%%%%%%

\maketitle

%\tableofcontents

\newpage
%%%%%%%%%%%%%%%%%%%%%%%%%%%%%%%%%%%%%%%%%%%%%%%%%%%%%%%%%%%%%
\section{Introduction}
%%%%%%%%%%%%%%%%%%%%%%%%%%%%%%%%%%%%%%%%%%%%%%%%%%%%%%%%%%%%%

Gauge theories are fundamental frameworks in modern physics for description of the interactions
in nature.
In a model where the gauge symmetry is spontaneously broken, the vector gauge field that acquires a mass
mediates a short-range force.
The massive vector field around a source charge drops off exponentially with the mass scale,
then the influence of the source charge by the massive vector field is limited
in a finite range of distance.
In other words, the source charge should be screened by some
appropriate configuration of the fields.

Motivated by color confinement, charge screening was investigated
in scalar electrodynamics \cite{Mandula1977,Adler_Piran,Bawin_Cugnon},
and Yang-Mills theories \cite{Sikivie_Weiss, William}.
It was reported that there exist minimum energy solutions which describe the screening of an external source
charge in gauge field models \cite{Mandula1977,Bawin_Cugnon, Adler_Piran,Sikivie_Weiss, William}.

In most of these works, singular shells are assumed as the source charge for convenience of analysis.
Smoothly extended charged objects with finite support are also possible sources to be screened.
As examples of the extended charged objects, we can consider non-topological solitons, which are studied
in coupled scalar fields systems \cite{Friedberg_Lee_Sirlin}, and a complex scalar field with non-trivial
self-interaction systems \cite{Coleman}.
Non-topological solitons of complex scalar fields coupled with a gauge field
are also investigated \cite{Lee_Stein-Schabes_Watkins_Widrow, Shi_Li, Gulamov_etal}.

We study, in this paper, screening of a smoothly extended source in
a system consisting of a U(1) gauge field and a complex Higgs scalar field with a potential
that causes spontaneous symmetry breaking. The purpose of this paper is
to clarify local configuration of the fields that screens the extended source, in detail.
We solve a coupled field equations numerically, and obtain spherically symmetric static solutions where
the charge screening occurs. We show that external charge is perfectly screened, that is,
the charge is canceled out everywhere by counter charge induced by the vector and scalar fields,
if the external charge spread widely compare to the mass scale of the vector field.

The organization of this paper is as follows.
In the next section, we present the basic system that is analyzed.
In section I\!I\!I, we reduce the system by assuming symmetry on the system, and set up external sources
and boundary conditions. Then, we obtain a set of ordinary differential equations to be solved.
In section I\!V, we perform numerical integrations of the equations in various cases for the external
sources, and show how the extended sources are screened.
Section V is devoted to summary and discussion.

%%%%%%%%%%%%%%%%%%%%%%%%%%%%%%%%%%%%%
\section{Basic System }
%%%%%%%%%%%%%%%%%%%%%%%%%%%%%%%%%%%%%
We consider an abelian Higgs system described by the Lagrangian density
 \begin{align}
  \mathcal{L}&=-(D_{\mu}\phi)^*(D^{\mu}\phi)-V(\phi)-\frac{1}{4}F_{\mu\nu}F^{\mu\nu},
  \label{eq:Lagrangian_0}
 \end{align}
where $F_{\mu\nu}:=\partial_{\mu}A_{\nu}-\partial_{\nu}A_{\mu}$ is the field strength of
a U(1) gauge field $A_{\mu}$, and
$\phi$ is a complex Higgs scalar field with the potential
\begin{align}
 	V(\phi)=\frac{\lambda}{4}(\phi^{\ast}\phi-\eta^2)^2,
 \label{eq:V}
\end{align}
where $\lambda$ and $\eta$ are positive constants.
The Higgs field $\phi$ couples to the gauge field by the covariant derivative %$D_{\mu}$
given by
\begin{align}
  D_{\mu}\phi &:=\partial_{\mu}\phi -ieA_{\mu}\phi,
 \label{eq:covariant derivative}
\end{align}
where $e$ is a coupling constant.
The Lagrangian density (\ref{eq:Lagrangian_0}) is invariant under local U(1) gauge transformations,
\begin{align}
  	\phi(x) &\to \phi'(x) =e^{i\chi(x)}\phi(x),
  \label{eq:gauge_tr_phi} \\
 	A_{\mu}(x) &\to A_{\mu}'(x) =A_{\mu}(x)+e^{-1}\partial_{\mu}\chi(x),
  \label{eq:gauge_tr_A}
\end{align}
where $\chi(x)$ is an arbitrary function.

The energy of the system is given by
\begin{align}
 	E&=\int d^3x \left(\left|D_{t}\phi\right|^2+(D_{i}\phi)^{\ast}(D^{i}\phi)
		+V(\phi)+\frac{1}{2}\left(E_iE^i+B_iB^i\right)\right) ,
\label{eq:energy}
\end{align}
where $E_i:=F_{i0}$,  $B^i:=1/2\epsilon^{ijk}F_{jk}$, and $i$ denotes spatial index.
In the vacuum state, which minimizes the energy \eqref{eq:energy},
$\phi$ and $A_{\mu}$ should take the form
 \begin{align}
  \phi = \eta e^{i\theta(x)}\ \text{and}\ A_{\mu}=e^{-1}\partial_{\mu}\theta,
  \label{eq:VEV}
 \end{align}
where $\theta$ is an arbitrary function.
Equivalently, eliminating $\theta$ we have
\begin{align}
 \phi^{\ast}\phi=\eta^2\ \text{and} \  D_{\mu}\phi=0.
 \label{eq:VEV2}
\end{align}
After the Higgs scalar field takes the vacuum expectation value $\eta$,
the gauge field $A_\mu$ absorbing the Nambu-Goldstone mode, the phase of $\phi$,
forms a massive vector field with the mass $m_A=\sqrt{2}e \eta$,
and the real scalar field that denotes a fluctuation of the amplitude of $\phi$
around $\eta$ acquires the mass $m_\phi=\sqrt{\lambda}\eta$.

In order to study the charge screening,
adding an extremal source current, $J^\mu$, coupled with $A_\mu$
to the original Lagrangian \eqref{eq:Lagrangian_0}, we consider the action\footnote{
The case of vanishing potential, $V(\phi)=0$, in which the symmetry does not break,
is studied in ref.\cite{Mandula1977},
and the case $V(\phi)=\frac{1}{2}m^2|\phi|^2$, in which partial screening occurs,
is studied in ref.\cite{Bawin_Cugnon}.
}
\begin{align}
	S =\int d^4x \left(-(D_{\mu}\phi)^*(D^{\mu}\phi)-\frac{\lambda}{4}(\phi^{\ast}\phi-\eta^2)^2
		-\frac{1}{4}F_{\mu\nu}F^{\mu\nu}-eA_{\mu}J^{\mu}\right).
  \label{eq:Lagrangian}
\end{align}
By varying (\ref{eq:Lagrangian}) with respect to $\phi^{\ast}$ and $A_{\mu}$,
we obtain the equations of motion
\begin{align}
  &D_{\mu}D^{\mu}\phi-\frac{\lambda}{2}\phi(\phi^{\ast}\phi-\eta^2)=0,
  \label{eq:V3}\\
 &\partial_{\mu}F^{\mu\nu}=e j_\text{ind}^\nu+eJ^{\nu},
  \label{eq:V4}
\end{align}
where $j_\text{ind}^\nu$ is the gauge invariant current density that consists
of $\phi$ and $A_\mu$ defined by
\begin{align}
   j_\text{ind}^\nu
	:= i\left(\phi^{\ast}(\partial^{\nu}-ieA^{\nu})\phi-\phi(\partial^{\nu}+ieA^{\nu})\phi^{\ast}\right) .
 \label{eq:scalar current}
\end{align}

%-----------------------------------------
\section{Spherically symmetric model}
%------------------------------------------
We consider a spherically symmetric and static external source in the form
\begin{align}
  eJ^t = \rho_\text{ext}(r), \quad \mbox{and}\quad eJ^i =0 ,
  \label{eq:V7}
\end{align}
where $t$ and $r$ are the time and the radial coordinates.
We also assume that the fields are spherically symmetric and stationary
in the form,
\begin{align}
  	\phi&=e^{i\omega t}f(r),
 \label{eq:V5}\\
	A_t&=A_t(r), \quad \text{and }\quad A_i=0,
  \label{eq:V6}
\end{align}
where $\omega$ is a constant, and $f(r)$ is a real function of $r$.
By using the gauge transformation \eqref{eq:gauge_tr_phi} and \eqref{eq:gauge_tr_A}
to incorporate the phase rotation of $\phi$, i.e., Nambu-Goldstone mode,
with $A_t$, we introduce a new variable $\alpha(r)$ as
\begin{align}
	\alpha(r):= A_t(r)-e^{-1}\omega.
\label{eq:alpha}
\end{align}
The charge density induced by the fields $\phi$ and $A_\mu$ defined
by \eqref{eq:scalar current} is written as
\begin{align}
	\rho_\text{ind} := e j^t_\text{ind} = -2e^2f^2\alpha.
\label{eq:rho_ind}
\end{align}

Substituting (\ref{eq:V7}) - (\ref{eq:alpha}) into (\ref{eq:V3}) and (\ref{eq:V4}), we obtain
\begin{align}
  &\frac{d^2f}{dr^2}+\frac{2}{r}\frac{df}{dr}+e^2f \alpha^2
	-\frac{\lambda}{2}f(f^2-\eta^2)=0,
\label{eq:eq_f}\\
  &\frac{d^2\alpha}{dr^2}+\frac{2}{r}\frac{d\alpha}{dr}
	+\rho_\text{ind}(r) +\rho_\text{ext}(r)=0.
\label{eq:eq_alpha}
\end{align}

Using the ansatz \eqref{eq:V5} and \eqref{eq:V6},
we rewrite the energy \eqref{eq:energy} for the symmetric system as
\begin{align}
	&E=4\pi \int_0^{\infty}r^2\epsilon(r)dr,
\label{eq:energy2}
\\
	&\epsilon := \epsilon_\text{Kin}+\epsilon_\text{Elast}
		+\epsilon_\text{Pot}+\epsilon_\text{ES},
\label{eq:total_energy_density}
\end{align}
where
\begin{align}
	&\epsilon_\text{Kin}:=\left|D_{t}\phi\right|^2=e^2f^2\alpha^2, \quad
	\epsilon_\text{Elast}:=(D_{i}\phi)^{\ast}(D^{i}\phi)=\biggl(\frac{df}{dr}\biggr)^2,
\label{eq:energy_density_1}
\\
	&\epsilon_\text{Pot}:=V(\phi)=\frac{\lambda}{4}(f^2-\eta)^2, \quad
\epsilon_\text{ES}:=\frac{1}{2}E_iE^i=\frac{1}{2}\biggl(\frac{d\alpha}{dr}\biggr)^2,
\label{eq:energy_density_2}
\end{align}
are density of kinetic energy, elastic energy, potential energy,
and electrostatic energy, respectively.

We consider a point source case, where the charge density is given by
the $\delta$-function as
\begin{align}
  	\rho_{\rm ext}(\vec{r})=q\delta^3(\vec{r}),
  \label{eq:point_source}
\end{align}
where $q$ denotes the total external charge,
and two extended source cases separately.
As the extended source cases, we discuss Gaussian distribution sources and
homogeneous ball sources. The both are smoothly distributed and have finite supports.

The charge density of the Gaussian distribution source is given by
\begin{align}
  	\rho_{\rm ext}(r)=\rho_0 \exp \left[-\biggl(\frac{r}{r_s}\biggr)^2\right] ,
\label{eq:Gaussian_source}
\end{align}
where $r_s$ is the width of the extended source.
The total external charge is assumed to be normalized as
\begin{align}
	4\pi \int_0^{\infty}r^2 \rho_\text{ext}(r)dr =q ,
\label{eq:normalize}
\end{align}
then the central density $\rho_0$ is given by
\begin{align}
	\rho_0=\frac{q}{\pi^{3/2}~ r_s^3}.
\label{eq:Gauss_center}
\end{align}
In the limit $r_s \to 0$, the charge density \eqref{eq:Gaussian_source} with \eqref{eq:Gauss_center}
reduces to the point source case \eqref{eq:point_source}.

The charge density of the homogeneous ball considered in this paper is given by
\begin{align}
  	\rho_\text{ext}(r)=\frac{\rho_0}{2} \left[ \tanh{\left(\frac{r_s-r}{\zeta_s}\right)}+1\right] ,
 \label{eq:Homogenious source}
\end{align}
where $r_s$ is the radius of the external source,
and $\zeta_s$ is the thickness of surface of the ball.
We assume $r_s \gg \zeta_s$ so that the charge density within the radius $r_s$ is
almost constant value $\rho_0$.
Then, the total external charge is $(4\pi/3) r_s^3\rho_0$ in this case.

We impose boundary conditions so that the fields are regular at the origin.
The regularity conditions for the spherically symmetric fields at the origin are
\begin{align}
  \frac{df}{dr}\to 0 \ , \ \frac{d\alpha}{dr}\to 0 \quad \mbox{as}\quad r\to 0.
\label{eq:BC_origin}
\end{align}
The energy density at the origin is finite for finite central values of $f$ and $\alpha$ .
At infinity, the fields should be in the vacuum state,
which minimizes $\epsilon$ in \eqref{eq:total_energy_density}.
Therefore, we impose the conditions
\begin{align}
  f \to \eta \ , \ \alpha \to 0 \quad \mbox{as}\quad r\to \infty .
\label{eq:BC_infty}
\end{align}

%----------------------------------------------t
\section{Numerical Calculations}
%----------------------------------------------

We use the relaxation method to obtain numerical solutions to the coupled
ordinary differential equations \eqref{eq:eq_f} and \eqref{eq:eq_alpha}.
In numerics, hereafter, we set $\eta=1$,
and scale the radial coordinate $r$ as $r \to \eta r$,
and scale the functions $f$, $\alpha$
as $f\to \eta^{-1}f$, $\alpha \to \eta^{-1}\alpha$, respectively.

%----------------------------------------------
\subsection{Point source}
%----------------------------------------------
Before the case of smoothly extended external charge distributions,
we consider the case of a point source \eqref{eq:point_source},
in which asymptotic behavior of the fields near the source is known analytically.
Main purposes of this subsection are confirmation of our numerical calculation
and observation of basic properties of the solutions.
We set $e=1/\sqrt{2}$ and $\lambda=1$ so that $r_{\phi}:=m_{\phi}^{-1}=1$
and $r_A:=m_A^{-1}=1$.

As is shown in Appendix \ref{app:A},
inspecting the equations \eqref{eq:eq_f} and \eqref{eq:eq_alpha},
we obtain the asymptotic behavior of $\alpha$ and $f$ near the origin are given by
\begin{align}
	\alpha(r)\sim \frac{q}{4\pi r},
\label{eq:q asymptotic origin}
\end{align}
and
\begin{align}
	f(r)\sim
\begin{cases}
	\quad b_1 r^{\beta}
  		 &\quad \text{for} \quad \kappa \leq \frac{1}{2}  ,\cr
	\quad\displaystyle
		\frac{b_1}{\sqrt{r}}\cos \left(\sigma \log r+ c_1\right)
 		&\quad \text{for}\quad \kappa> \frac{1}{2} ,
\end{cases}
\label{eq:f asymptotic origin}
\end{align}
where
$\kappa:=eq/4\pi$, $\beta:=\frac{1}{2}\left(-1+\sqrt{1-4\kappa^2}\right)$,
$\sigma:=\frac{1}{2}\sqrt{4\kappa^2-1}$, and $b_1, c_1$ are constants.
We should note that the behavior of $f(r)$ critically depends on the parameter $\kappa$.

%%%%%%%%%%%%%%%%%%%%%%%%%%%%%%%%%%%%%%%%%%%%%%%%%%%%%%%%%%%%%%%%%%%%%%%
\begin{figure}[H]
\includegraphics[width=7cm]{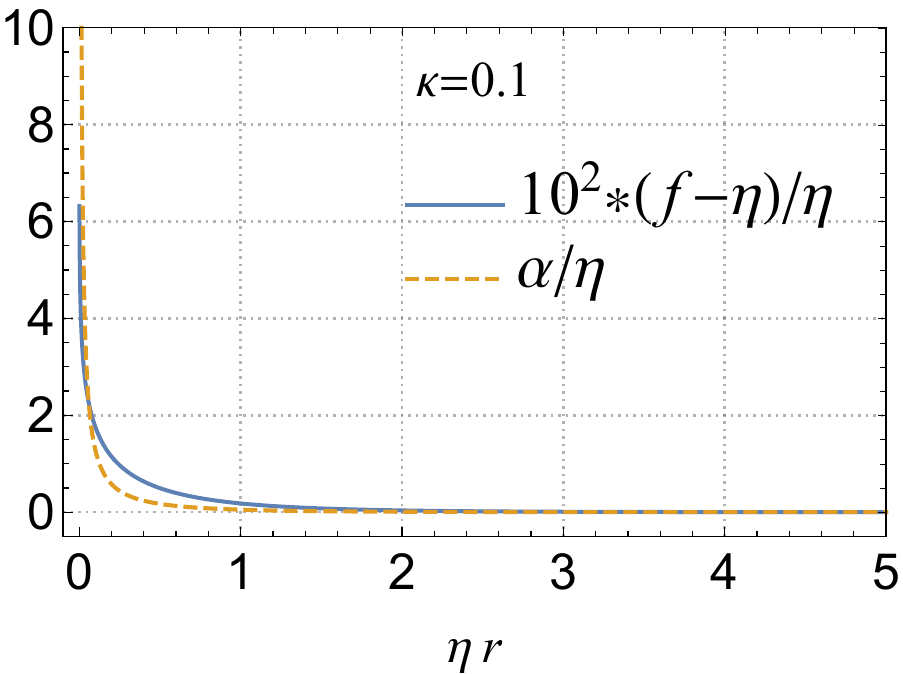} \hspace{1cm}
\includegraphics[width=7cm]{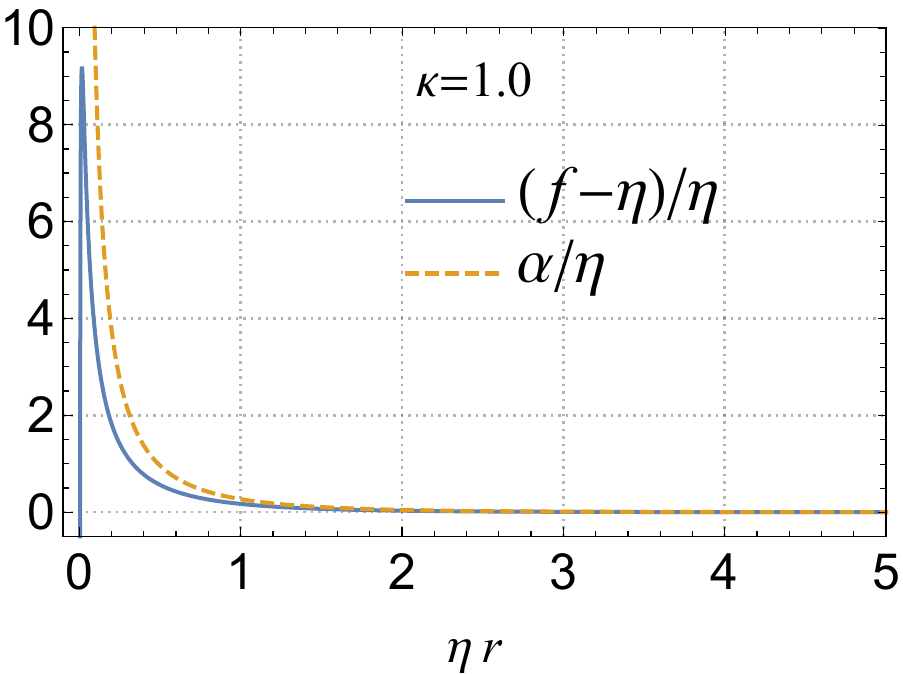}
\\
\centering
\begin{minipage}{0.9\hsize}
\caption{
Numerical solutions of $f(r)$ and $\alpha(r)$ for a point source
in the case $\kappa=0.1$ (left panel),
and in the case $\kappa=1.0$ (right panel).
\label{fig:deltasolution}
}
\end{minipage}
\end{figure}
%%%%%%%%%%%%%%%%%%%%%%%%%%%%%%%%%%%%%%%%%%%%%%%%%%%%%%%%%%%%%%%%%%%%%%%

%%%%%%%%%%%%%%%%%%%%%%%%%%%%%%%%%%%%%%%%%%%%%%%%%%%%%%%%%%%%%%%%%%%%%%%
\begin{figure}[H]
\begin{minipage}{\hsize}
\centering
\includegraphics[width=7.9cm]{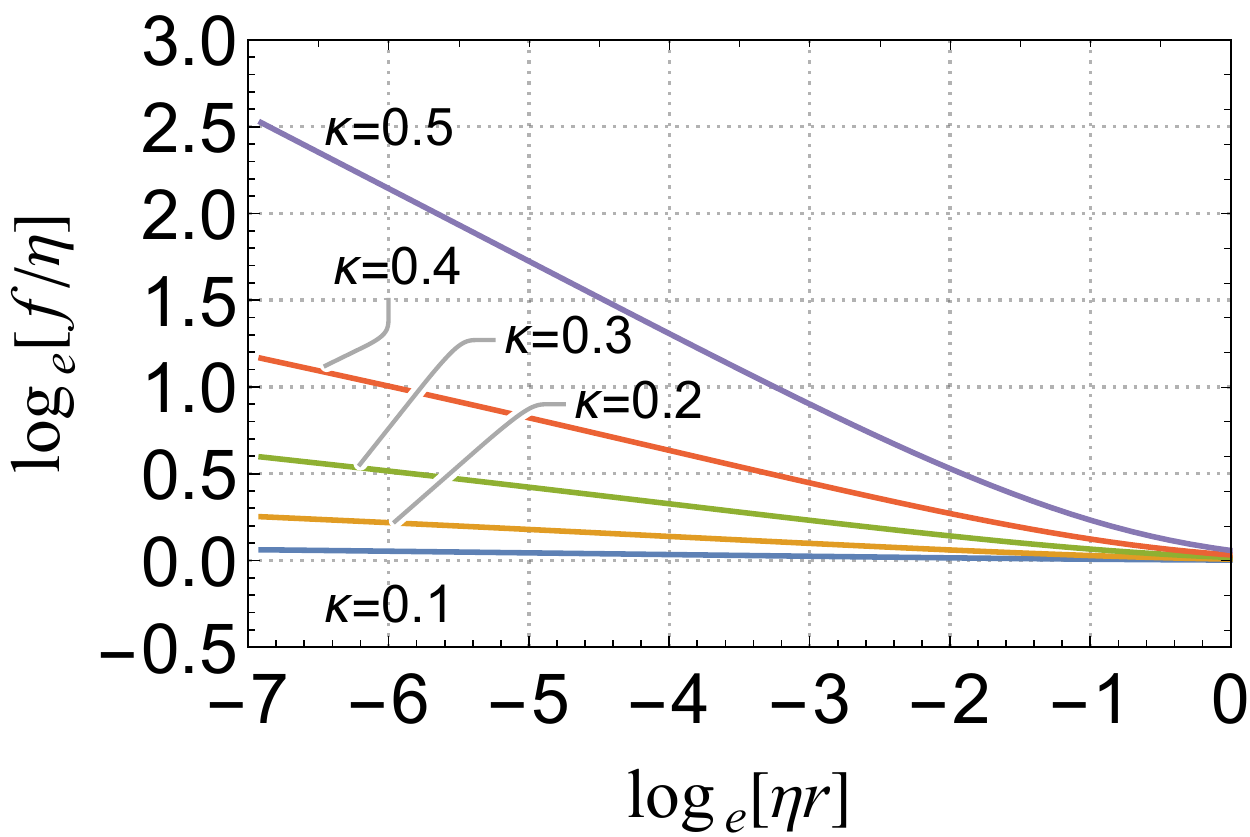}~~
\includegraphics[width=8.1cm]{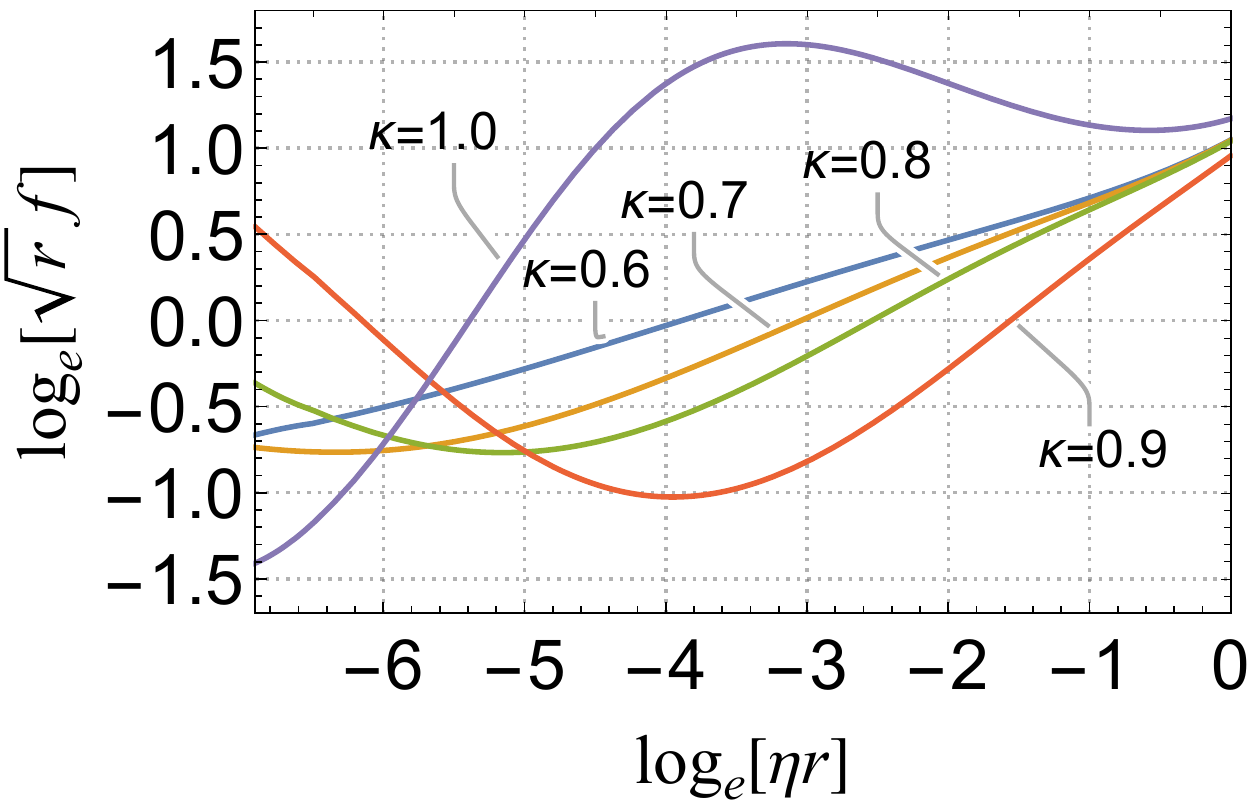}
\end{minipage}
\begin{minipage}{0.06\hsize}
        \vspace{10mm}
\end{minipage}
\\
\begin{minipage}{0.5\hsize}
\centering
\includegraphics[width=7.8cm]{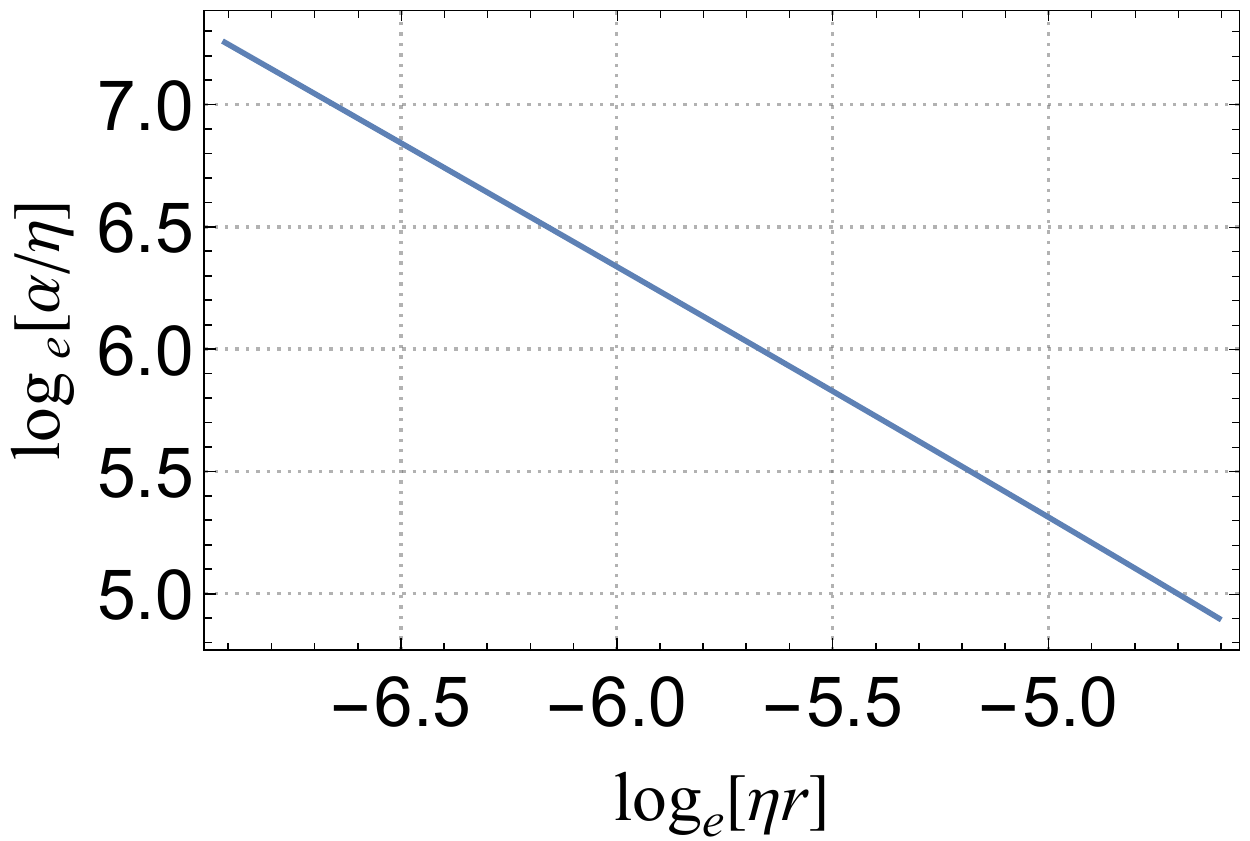}
\end{minipage}~~~~
\begin{minipage}{0.5\hsize}
\caption{
Asymptotic behaviors of the function $f(r)$ and $\alpha(r)$ near the origin.
Behaviors of $f(r)$ for $\kappa=0.1 - 0.5$ (left upper panel)
and for $\kappa=0.6 - 1.0$ (right upper panel) are shown.
Behaviors of $\alpha(r)$ (lower panel) are the same for $\kappa=0.1 - 1.0$.
\label{fig:Gaussianconfiguration3}
}
\end{minipage}

\end{figure}
%%%%%%%%%%%%%%%%%%%%%%%%%%%%%%%%%%%%%%%%%%%%%%%%%%%%%%%%%%%%%%%%%%%%%%%

On the other hand, the asymptotic behaviors at infinity are given by
\begin{align}
  \alpha(r)&\sim\frac{a_2}{r}\exp\left(-\frac{r}{r_{A}}\right),
\label{eq:q asymptotic inf2}
\\
	f(r) &\sim \eta +\frac{b_2}{r}\exp\left(-\frac{r}{r_{\phi}}\right),
\label{eq:f asymptotic inf2}
 \end{align}
where  $a_2$ and $b_2$ are constants.

\begin{figure}[H]
\includegraphics[width=8cm]{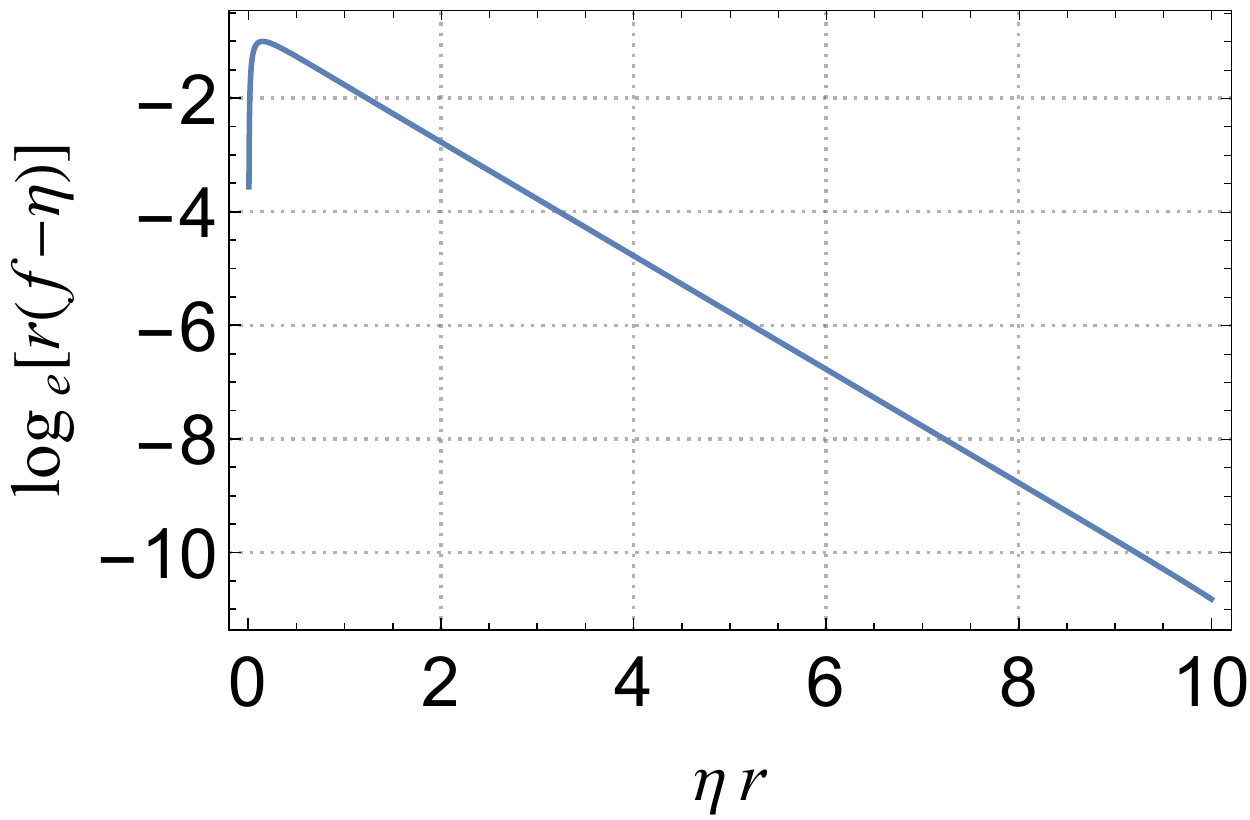}~~
\includegraphics[width=8cm]{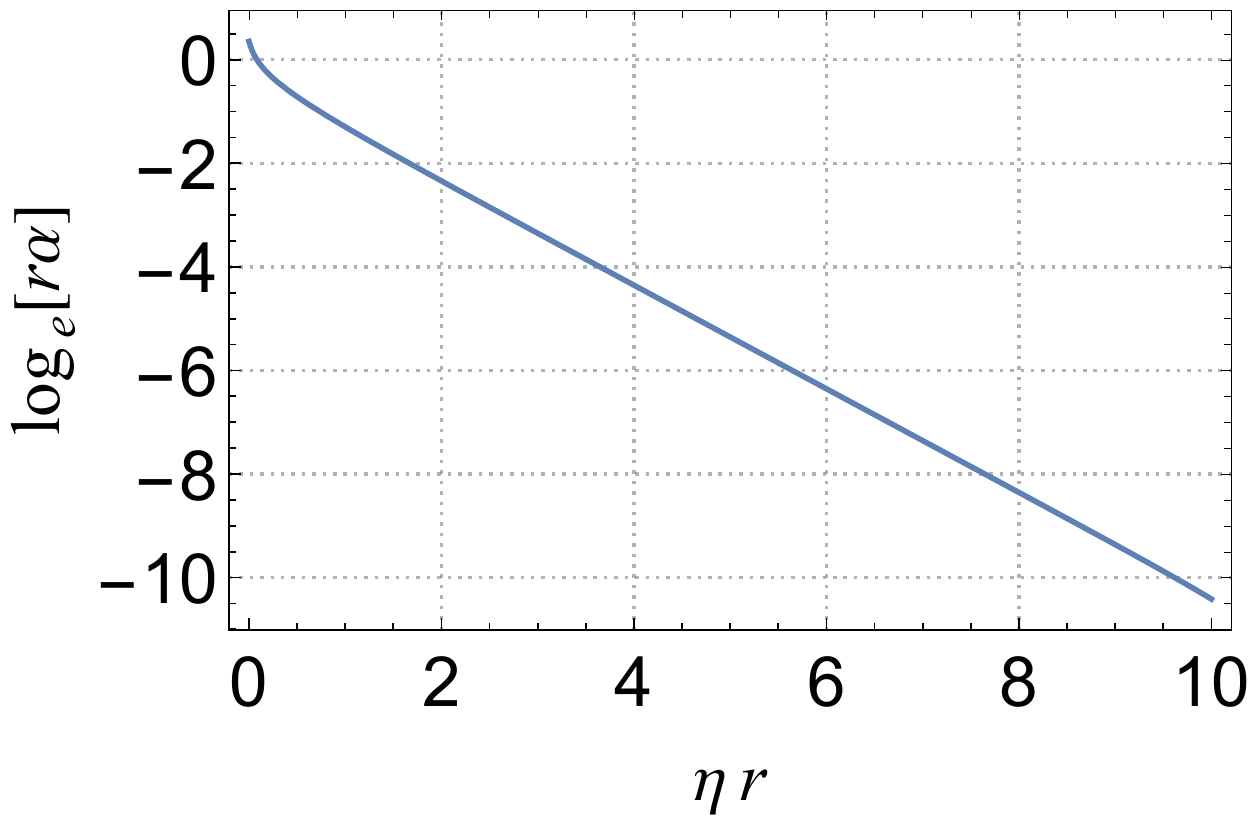}
\centering
\begin{minipage}{0.9\hsize}
\caption{
Asymptoric behaviors of the function $f(r)$ and $\alpha(r)$ at infinity.
The functions $r(f(r)-\eta)$ (left panel) and $r \alpha(r)$ (right panel) are shown
in logarithmic scale.
\label{fig:Gaussianconfiguration2}
}
\end{minipage}
\end{figure}

Here, we solve equations \eqref{eq:eq_f} and \eqref{eq:eq_alpha} numerically,
and study basic properties of the system.
Typical behaviors of the functions $f(r)$ and $\alpha(r)$
are shown in Fig.\ref{fig:deltasolution}.
Especially, the behaviors of $f$ and $\alpha$ near the origin are shown
in Fig.\ref{fig:Gaussianconfiguration3}.
In the case of $\kappa \leq 1/2$, $f$ is given by the power function of $r$,
while in the case of $\kappa > 1/2$, oscillatory behaviors appear.
The function $\alpha$ is in proportion to $r^{-1}$ independent with $\kappa$.
These behaviors coincide with \eqref{eq:q asymptotic origin} and
\eqref{eq:f asymptotic origin}.
The asymptotic behaviors of the functions $f$ and $\alpha$ in a distant region coincide with
\eqref{eq:q asymptotic inf2} and \eqref{eq:f asymptotic inf2} as
shown in Fig.\ref{fig:Gaussianconfiguration2}.

%%%%%%%%%%%%%%%%%%%%%%%%%%%%%%%%%%%%%%%%%%%%%%%%%%%%%%%%%%%%%%%%%%%%%%%
%%%%%%%%%%%%%%%%%%%%%%%%%%%%%%%%%%%%%%%%%%%%%%%%%%%%%%%%%%%%%%%%%%%%%%%

\begin{figure}[H]
\includegraphics[width=8cm]{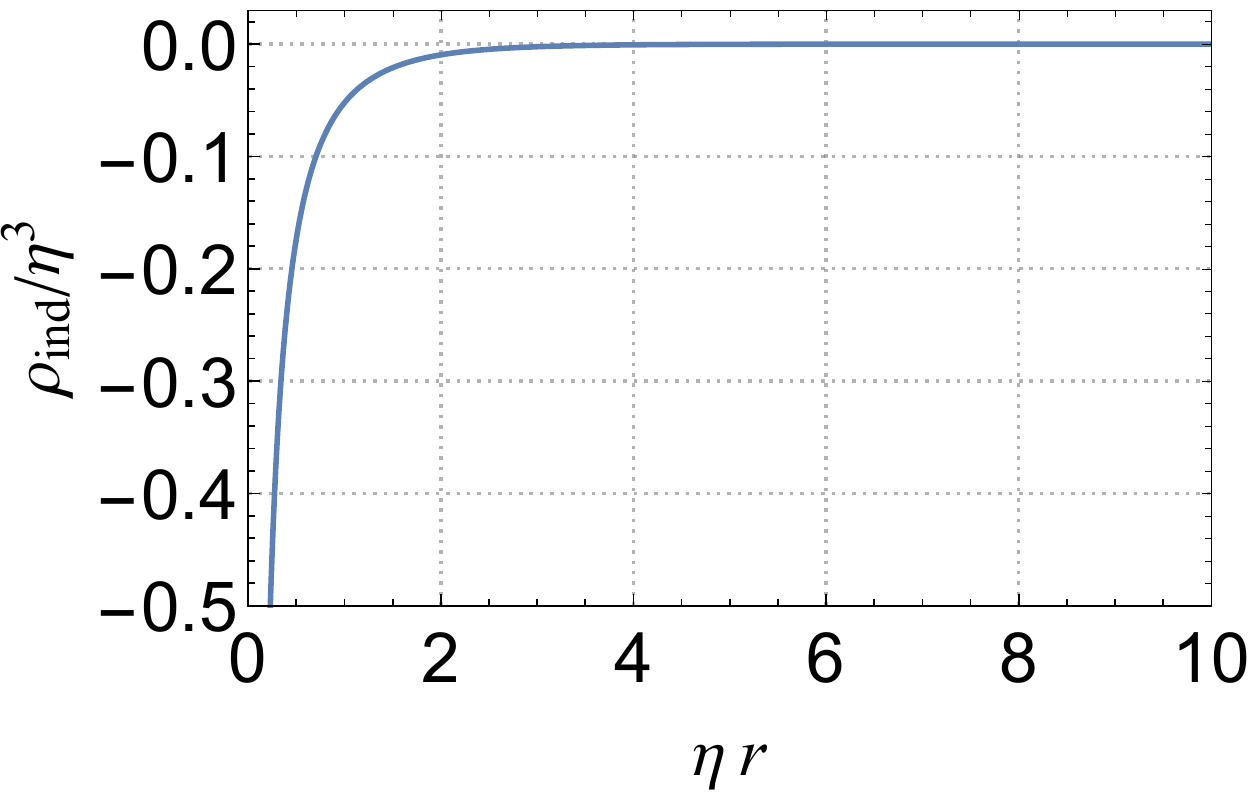}
~
\includegraphics[width=7.8cm]{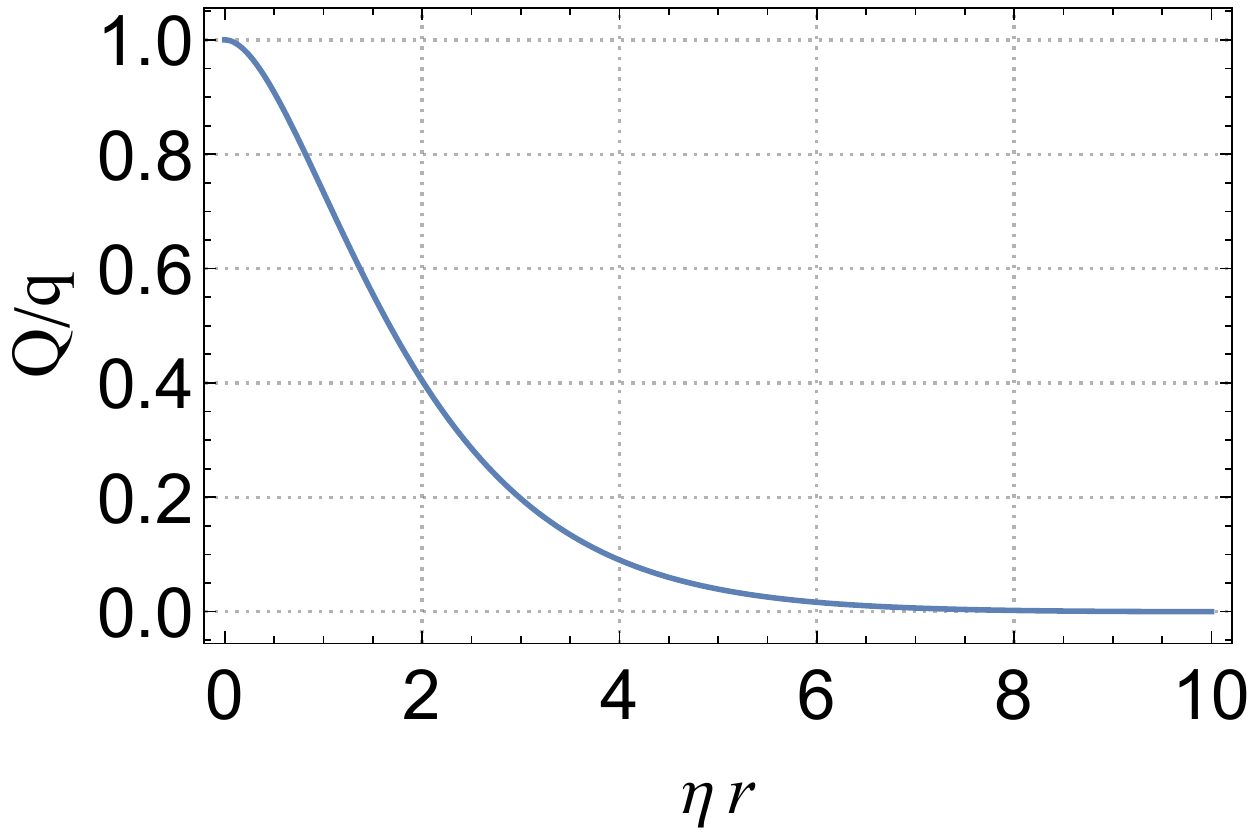}
\\
\centering
\begin{minipage}{0.9\hsize}
\caption{
The induced charge density, $\rho_\text{ind}$, (left pannel),
and the total charge within radius $r$, $Q(r)$, (right pannel)
are plotted for the case of point charge source.
\label{fig:deltachargecomponent}
}
\end{minipage}
\end{figure}

The induced charge density $\rho_\text{ind}$ in \eqref{eq:rho_ind} is plotted
in the left panel of Fig.\ref{fig:deltachargecomponent} as a function of $r$.
The induced charge, whose sign is opposite to the external source charge,
distributes as a cloud around the point charge source.
We define the total charge within the radius $r$, say $Q(r)$, by
\begin{align}
  Q(r):&=4\pi \int_0^{r}\tilde{r}^2 \rho_\text{total}(\tilde{r})d\tilde{r},
\label{eq:total charge}
\end{align}
where the total charge density $\rho_\text{total}(r)$ is defined by
\begin{align}
	\rho_\text{total}(r):= \rho_\text{ext}(r)+\rho_\text{ind}(r).
\end{align}
As shown in Fig.\ref{fig:deltachargecomponent}, $Q(r)$ is monotonically
decreasing function of $r$. It means that the positive charge of
the external source is screened by the induced negative charge cloud.
In the region near the point source the charge is partly screened, i.e.,
$1 > Q(r)/q >0$  and at a large distance the charge is totally screened,
i.e., $Q(r)/q=0$.

%%%%%%%%%%%%%%%%%%%%%%%%%%%%%%%%%%%%%%%%%%%%%%%%%%%%%%%%%%%%%%%%%%%%%%%
%%%%%%%%%%%%%%%%%%%%%%%%%%%%%%%%%%%%%%%%%%%%%%%%%%%%%%%%%%%%%%%%%%%%%%%

\begin{figure}[H]
\centering
\includegraphics[width=8cm]{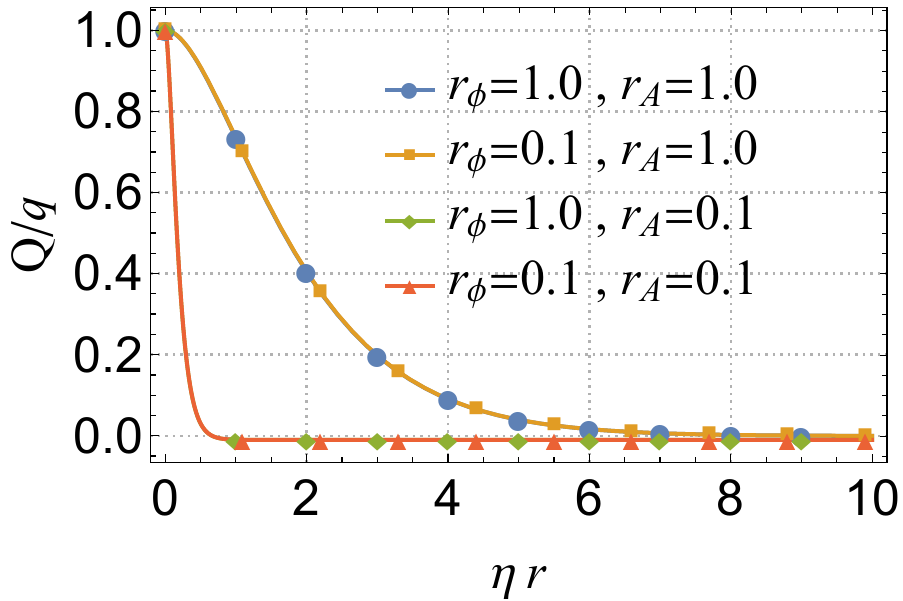}
\caption{
The total charge $Q(r)$ for some sets of parameters $(r_{\phi}, r_A)$
and fixed $q = 0.1$.
}
\label{fig:screeningratiochange}
\end{figure}

For some sets of two characteristic length scales $(r_\phi, r_A)$,
the function $Q(r)$ is plotted in Fig.\ref{fig:screeningratiochange}.
We see that the shape of $Q(r)$ does not depend on $r_\phi$,
while the width of $Q(r)$ is given by $r_A$.
In any case, in a distant region where $r \gg r_A$, charge is totally screened.
Except the neighborhood of the origin, as shown in Fig.\ref{fig:deltasolution},
$f$ takes the vacuum expectation value $\eta$.
The massive gauge mode with mass $m_A$
causes the charge screening, with the size of $r_A=m_A^{-1}$.

%%%%%%%%%%%%%%%%%%%%%%%%%%%%%%%%%%%%%%%%%%%%%%%%%%%%%%%%%%%%%%%%%%%
%%%%%%%%%%%%%%%%%%%%%%%%%%%%%%%%%%%%%%%%%%%%%%%%%%%%%%%%%%%%%%%%%%%
%

%----------------------------------------------
\subsection{Gaussian distribution source}
%----------------------------------------------
For the first example of smoothly extended source,
we consider the external charge density given by
the Gaussian distribution \eqref{eq:Gaussian_source}.
As the boundary conditions, we impose the regularity conditions \eqref{eq:BC_origin} at the origin,
and the vacuum condition \eqref{eq:BC_infty} at infinity.

For the extended external sources,
the behaviors of $f$ and $\alpha$ do not depend critically on the value of $\kappa$
unlike the point source case. So, we concentrate on the case $\kappa=e/(4\pi)$, i.e., $q=1$.
We fix $r_{\phi}=1$, $r_A=1$,
and perform numerical calculation with several values of $r_s$ that denotes the thickness
of the external source.

%---------------------------------------
\subsubsection{Field configurations}
%---------------------------------------

\begin{figure}[H]
\centering
\includegraphics[width=7cm]{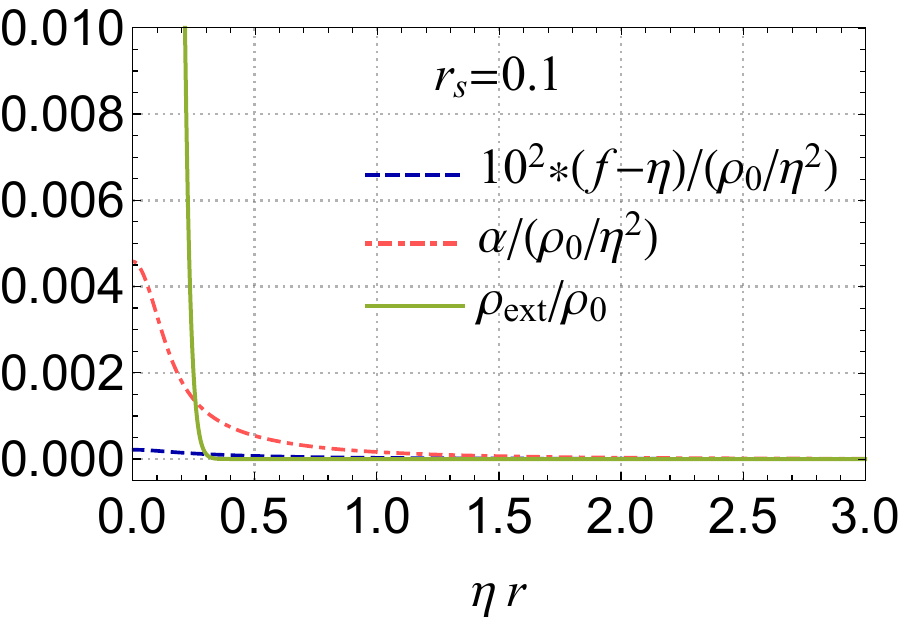}~~
%\centering
\includegraphics[width=6.8cm]{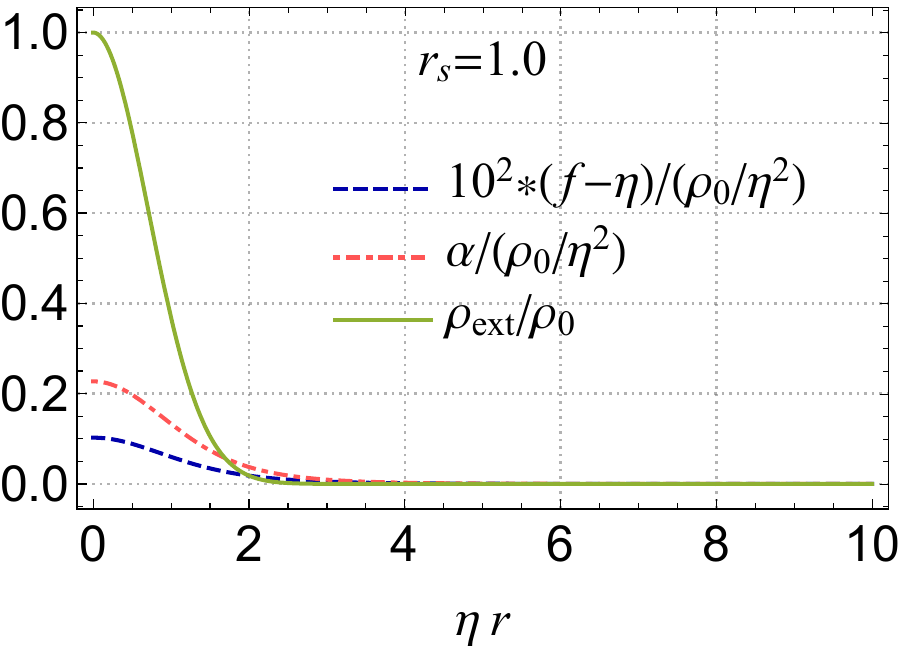}
\\
\begin{minipage}{0.06\hsize}
        \vspace{5mm}
\end{minipage} \\
\centering
\includegraphics[width=6.8cm]{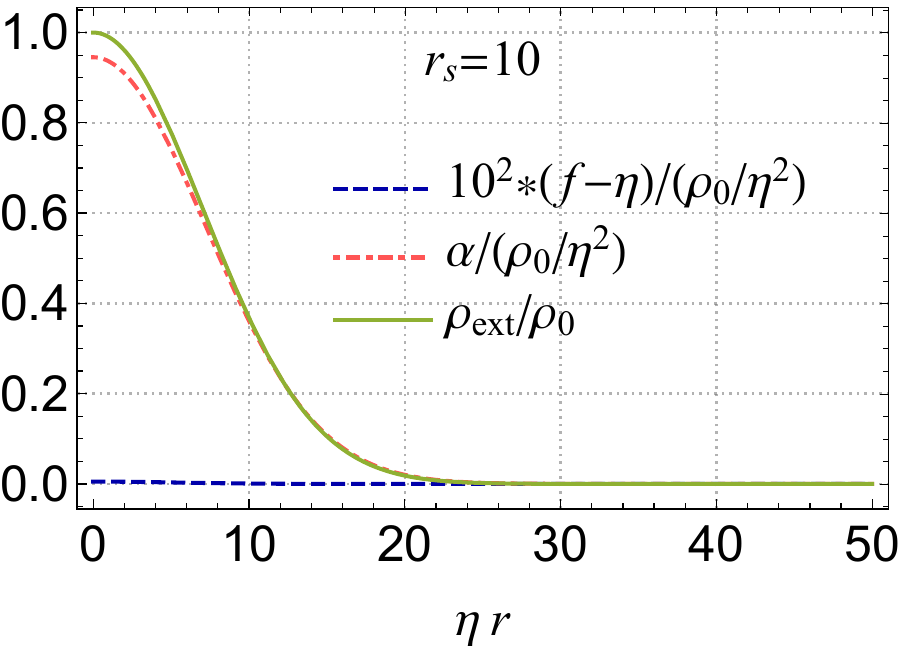}~~
%\centering
\includegraphics[width=6.8cm]{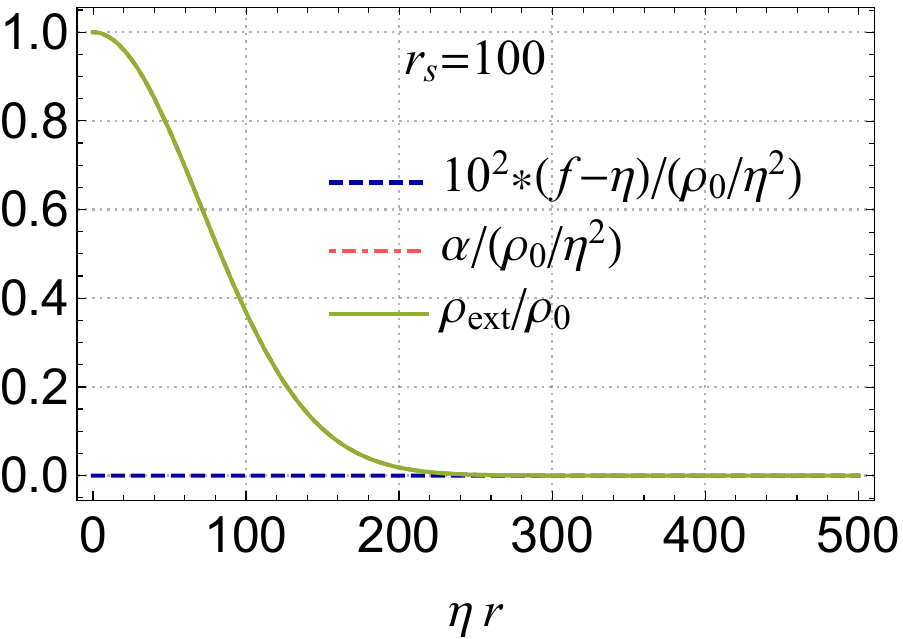}
\begin{minipage}[b]{0.9\hsize}
\caption{
Numerical solutions in the case of Gaussian distribution sources.
Behaviors of $f(r)$ and $\alpha(r)$
for $r_s=0.1, 1, 10, 100$ are drawn together with $\rho_\text{ext}$ as functions of $r$.
In the case of $r_s=100$, $\alpha(r)$ coincides with $\rho_\text{ext}$ (see the lower right panel).
\label{fig:Gaussian_configuration}
}
\end{minipage}
\end{figure}

\begin{figure}[H]
\includegraphics[width=7cm]{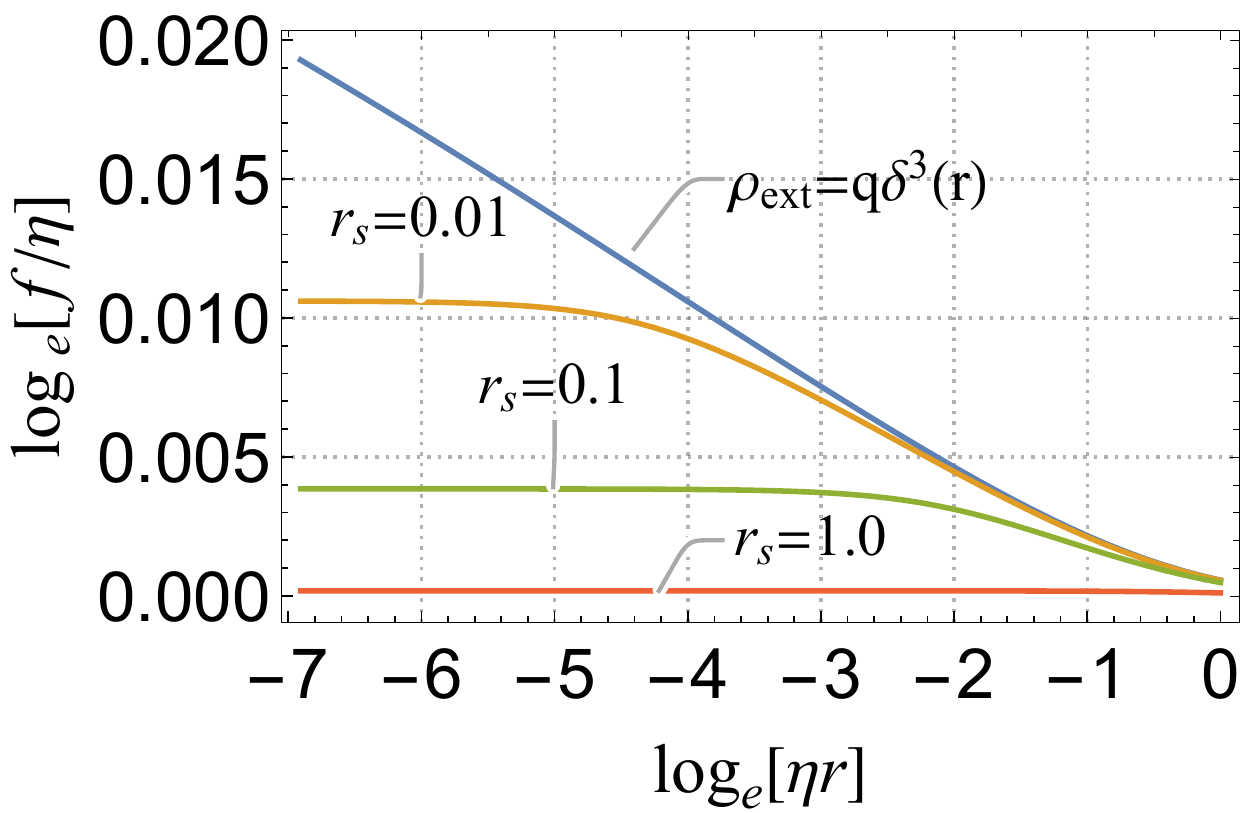}~~
\includegraphics[width=6.5cm]{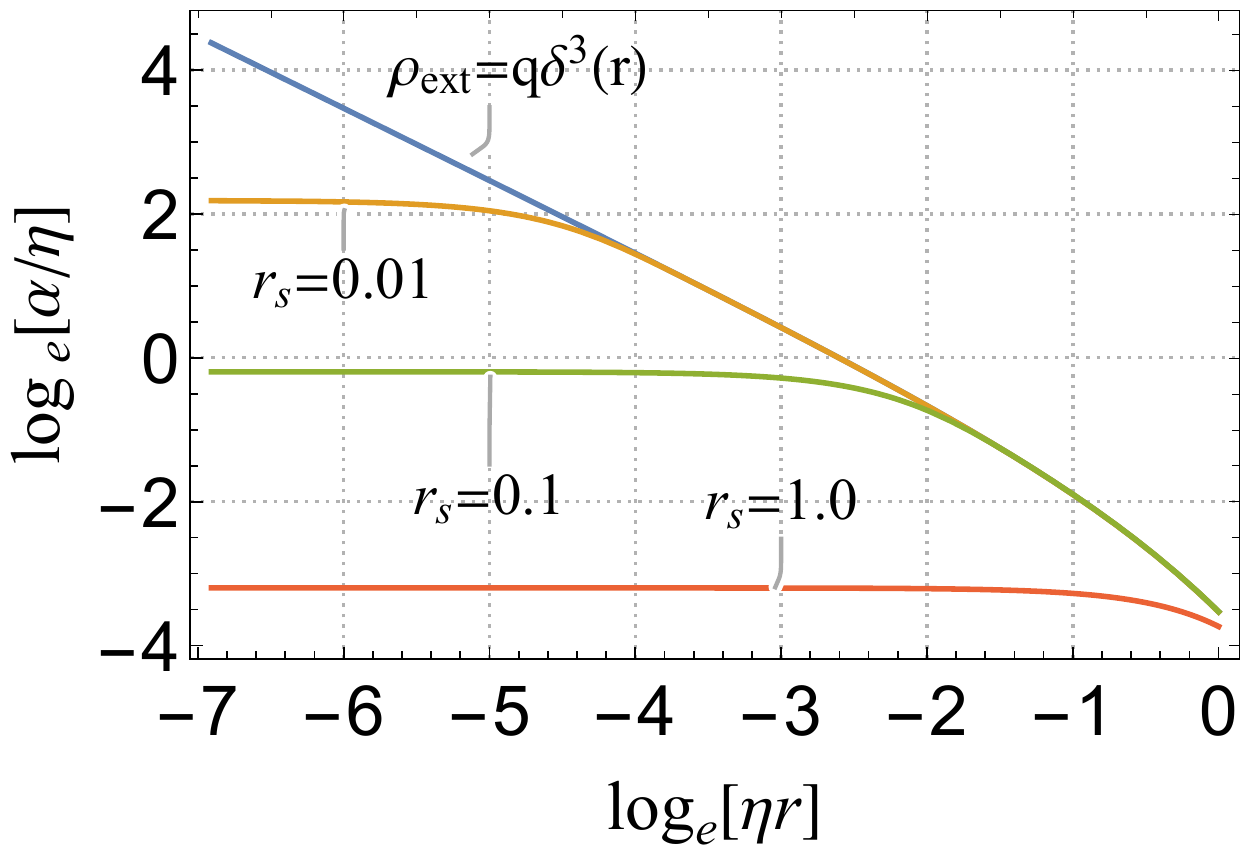}
\centering
\begin{minipage}[b]{0.9\hsize}
\caption{
Behaviors of $f(r)$ and $\alpha(r)$ for various $r_s$.
As $r_s$ decreases to zero, the configurations of $f$ and $\alpha$ approach
to the ones in the point source case.
\label{fig:xicharge_configuration}
}
\end{minipage}
\end{figure}

By numerical calculations, we show typical behaviors of the function $f$ and $\alpha$
with the external charge density $\rho_\text{ext}$
in the cases of
$r_s=0.1, 1, 10$, and $100$ in Fig.\ref{fig:Gaussian_configuration}.
We see that the function $f$ and $\alpha$ change in their shapes with $r_s$.
Especially, for the thin source case, $r_s\ll r_A$, numerical solutions are shown
in Fig.\ref{fig:xicharge_configuration}.
As $r_s$ approaches to zero, since the normalized Gaussian function \eqref{eq:Gaussian_source}
with \eqref{eq:Gauss_center} reduces to the $\delta$-function \eqref{eq:point_source},
then as we expected, the function $f$ and $\alpha$ approach to the solutions
for the point source case discussed in the previous subsection.
In the thick source case, $r_s  > r_A$, the widths of $f$ and $\alpha$ are order of $r_s$.
Typical behaviors can be understood by analytical method given in Appendix \ref{app:approx_sol_Gaussian}.

%%%%%%%%%%%%%%%%%%%%%%%%%%%%%%%%%%%%%%%%%%%%%%%%%%%%%%%%%%%%%%%%%%
%%%%%%%%%%%%%%%%%%%%%%%%%%%%%%%%%%%%%%%%%%%%%%%%%%%%%%%%%%%%%%%%%%

%%%%%%%%%%%%%%%%%%%%%%%%%%%%%%%%%%%%%%%%%%%%%%%%%%%%%%%%%%%%%%%%%%
%%%%%%%%%%%%%%%%%%%%%%%%%%%%%%%%%%%%%%%%%%%%%%%%%%%%%%%%%%%%%%%%%%

%---------------------------------------
\subsubsection{Charge screening}
\begin{figure}[H]
\centering
\includegraphics[width=6.5cm]{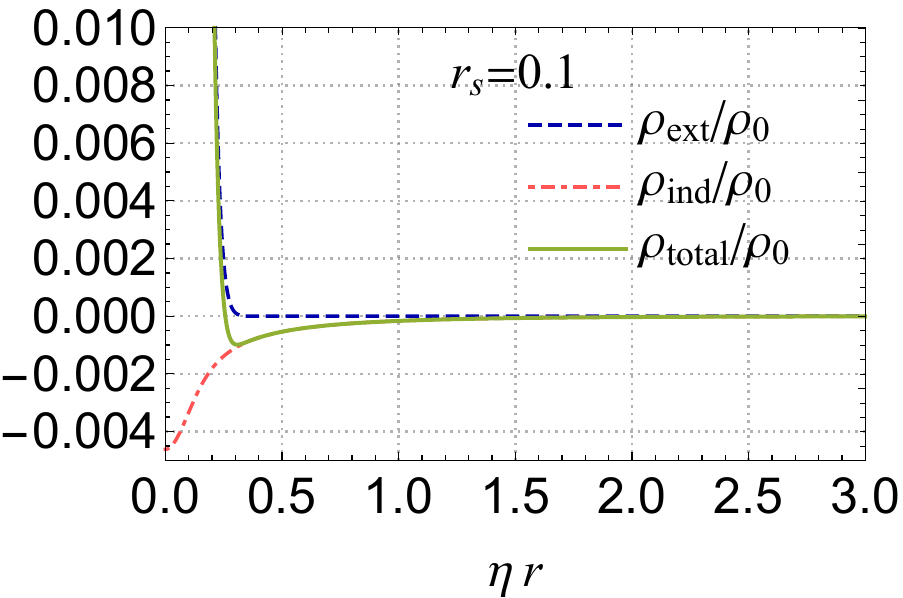}~~
%\centering
\includegraphics[width=6.2cm]{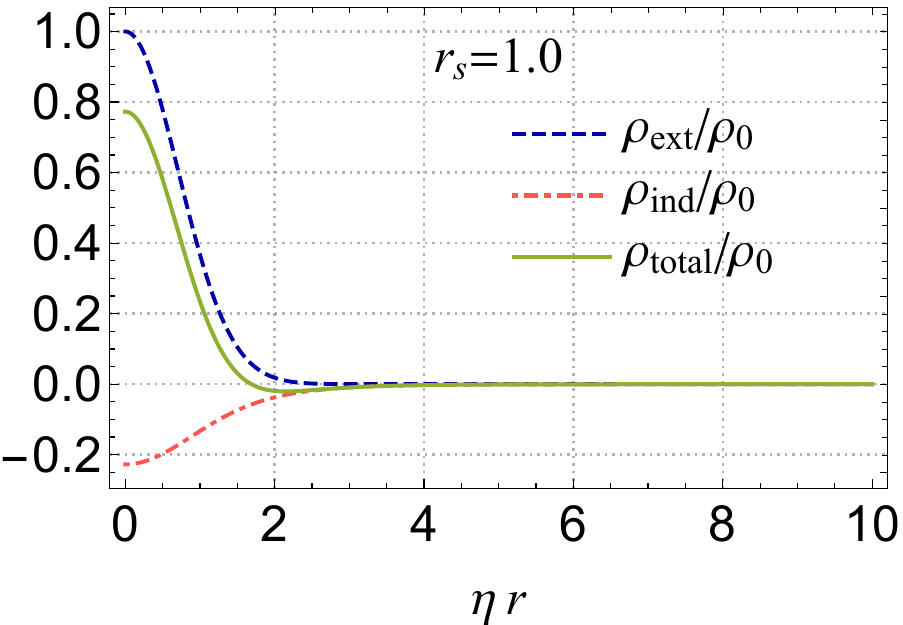}
\begin{minipage}{0.06\hsize}
        \vspace{5mm}
\end{minipage}
\\
\centering
\includegraphics[width=6.2cm]{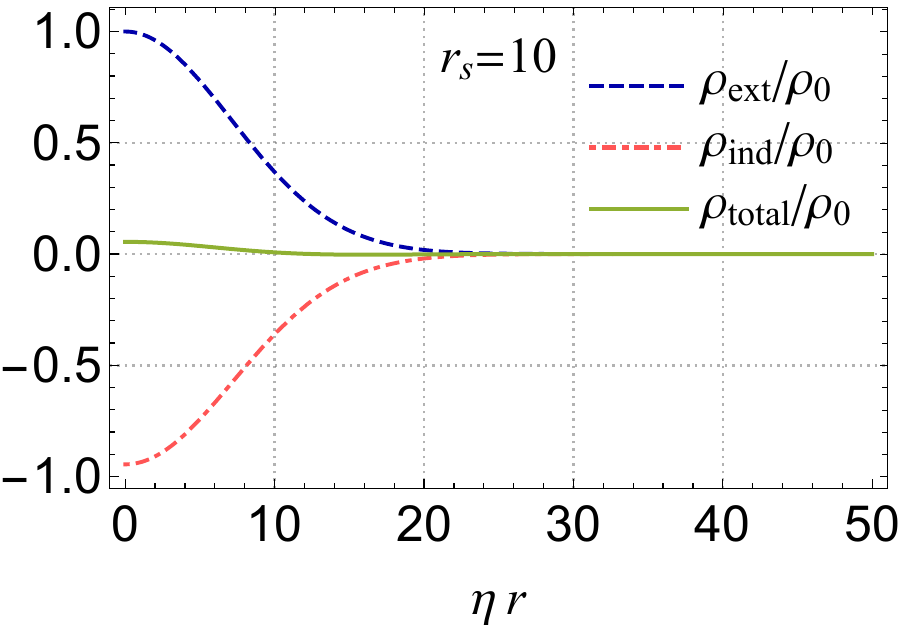}~~
%\centering
\includegraphics[width=6.2cm]{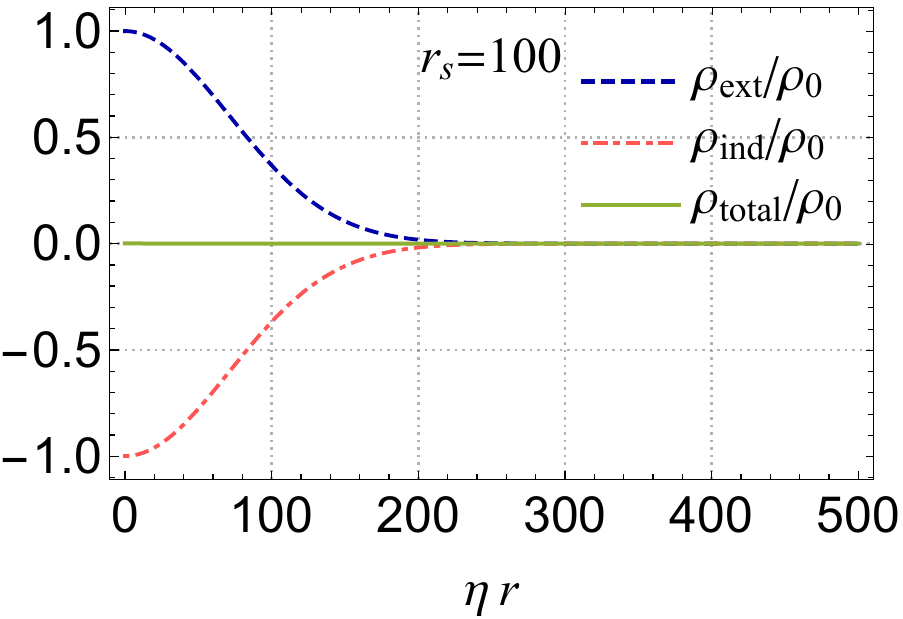}
\begin{minipage}[b]{0.9\hsize}
\caption{
The external charge density, $\rho_\text{ext}(r)$,
the induced charge density, $\rho_\text{ind}(r)$, and sum of them, $\rho_\text{total}(r)$,
are plotted as functions of $r$ for $r_s= 0.1, 1, 10, 100$.
\label{fig:Gaussianchargedensity}
}
\end{minipage}
\end{figure}

\begin{figure}[H]
\centering
\includegraphics[width=8cm]{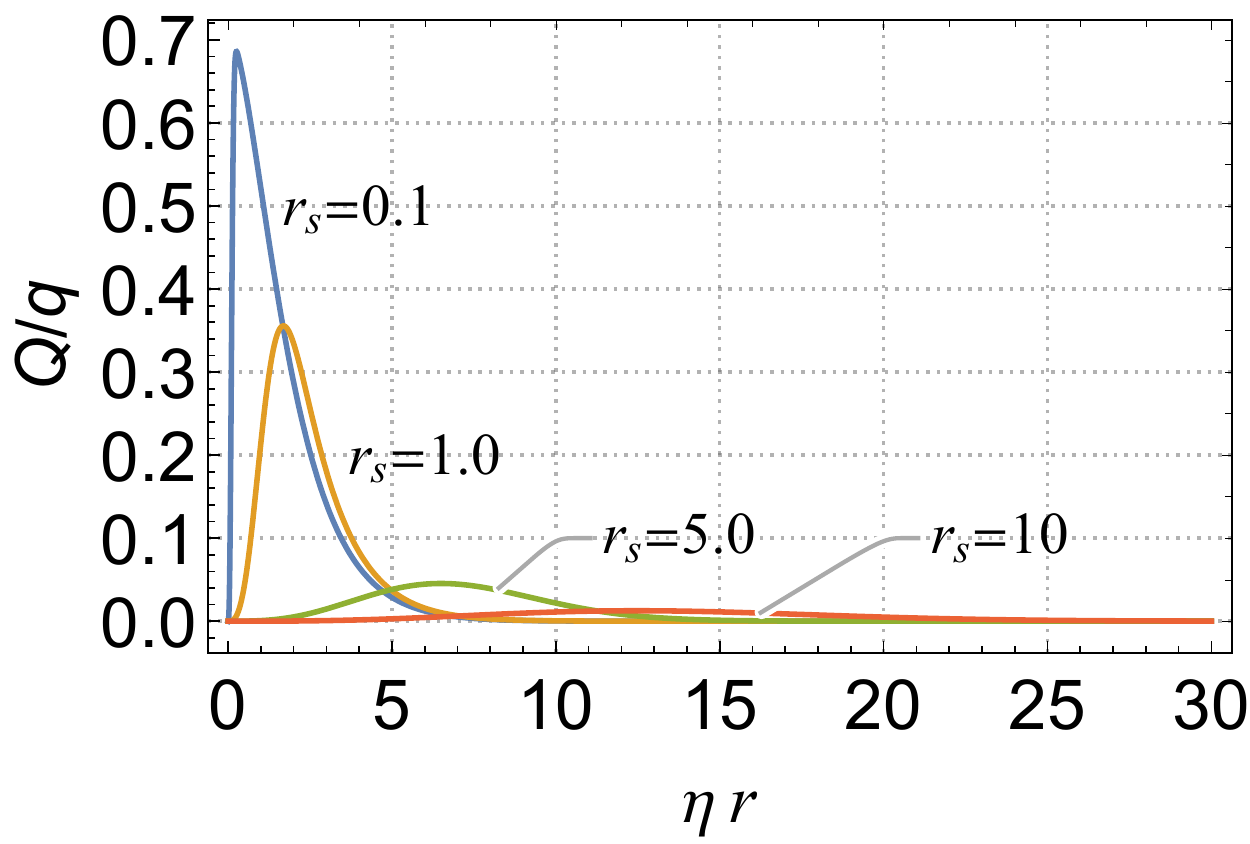}
\caption{
The total charges within radius $r$, $Q(r)$, are plotted for  $r_s= 0.1, 1, 5, 10$.
\label{fig:Gaussiantotalcharge}
}
\end{figure}

%---------------------------------------
We depict the induced charge density $\rho_\text{ind}(r)$ with the external charge
density $\rho_\text{ext}(r)$ in Fig.\ref{fig:Gaussianchargedensity}.
The sign of $\rho_\text{ind}$ is opposite to $\rho_\text{ext}$.
In the central region of $r_s=0.1$ and $r_s=1$ cases, we find that $|\rho_\text{ext}|$ is larger than
$|\rho_\text{ind}|$, i.e., total charge density $\rho_\text{total}(r)$
has the same sign with $\rho_\text{ext}(r)$.
As $r$ increases, $|\rho_\text{ind}|$ exceeds $|\rho_\text{ext}|$.
In the region $r \gg r_A$, the both $\rho_\text{ext}$ and $\rho_\text{ind}$
decrease quickly to zero.
As shown in Fig.\ref{fig:Gaussiantotalcharge},  $Q(r)$, the total charge within radius $r$,
decreases to zero in the region $r \gg {\rm max}(r_A, r_s)$,
it means the external charge is totally screened by the induced charge cloud
for a distant observer.

%%%%%%%%%%%%%%%%%%%%%%%%%%%%%%%%%%%%%%%%%%%%%%%%%%%%%%%%%%%%%%%%%%%%%%
%%%%%%%%%%%%%%%%%%%%%%%%%%%%%%%%%%%%%%%%%%%%%%%%%%%%%%%%%%%%%%%%%%%%%%

In the case of $r_s\ll r_A$, the width of the induces charge cloud is the order of
$r_A$, while in the case of $r_s \geq r_A$, the width is almost same as $r_s$.
In the case of $r_s =100$, we have
\begin{align}
	\rho_\text{ind}(r) = -\rho_\text{ext}(r),
\end{align}
as is justified by \eqref{eq:thick_alpha}.
Then, $\rho_\text{total}$ vanishes everywhere, equivalently $Q(r)$ vanishes everywhere.
We call this \lq perfect screening\rq .

%%%%%%%%%%%%%%%%%%%%%%%%%%%%%%%%%%%%%%%%%%%%%%%%%%%%%%%%%%%%%%%%%%%%%%
%%%%%%%%%%%%%%%%%%%%%%%%%%%%%%%%%%%%%%%%%%%%%%%%%%%%%%%%%%%%%%%%%%%%%%

%---------------------------------------
\subsubsection{Energy of the cloud}
\begin{figure}[H]
\begin{center}
\includegraphics[width=7cm]{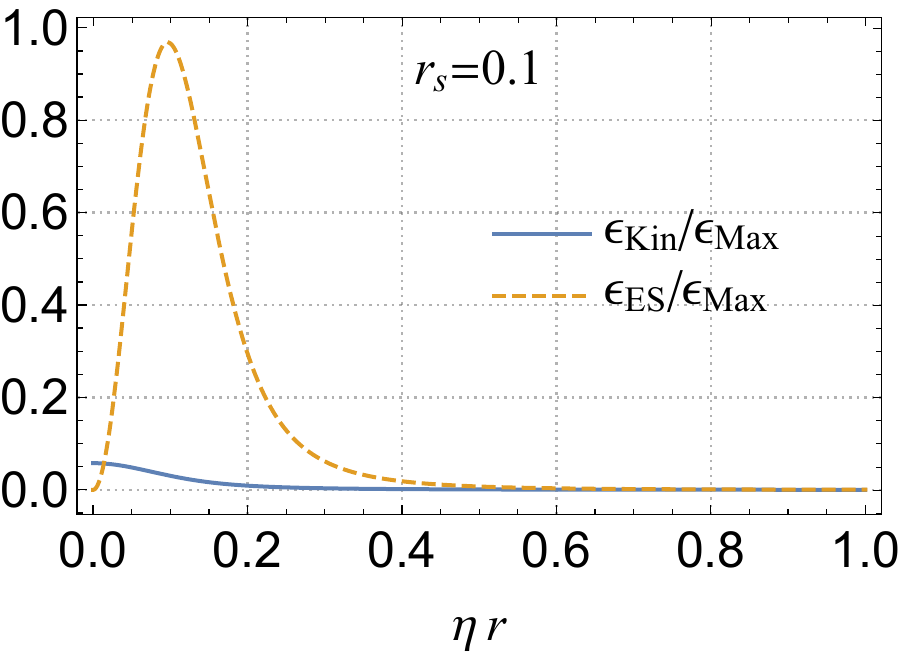}~~
\includegraphics[width=7cm]{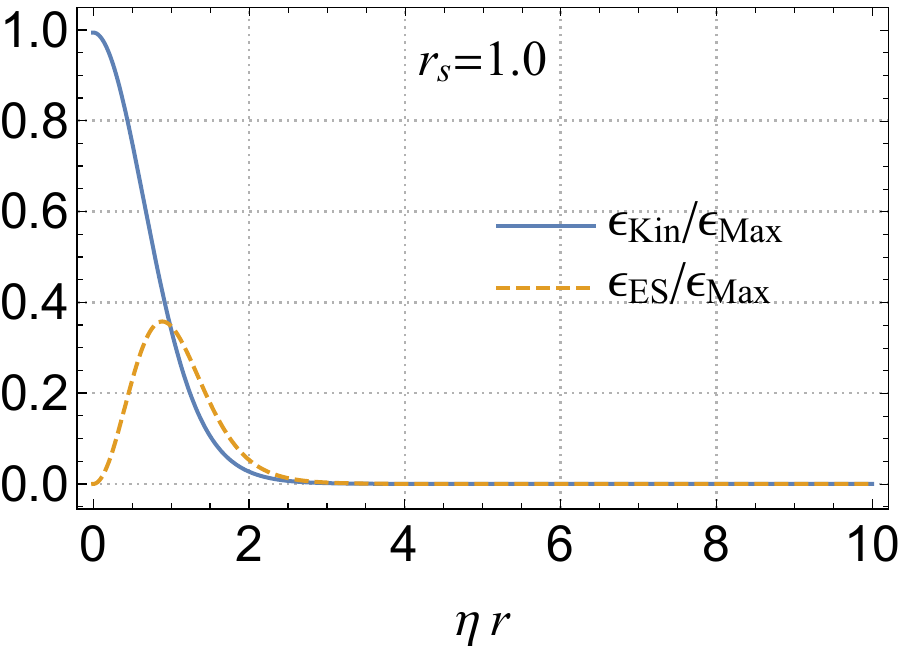}
\\
\vspace{1cm}
\includegraphics[width=7cm]{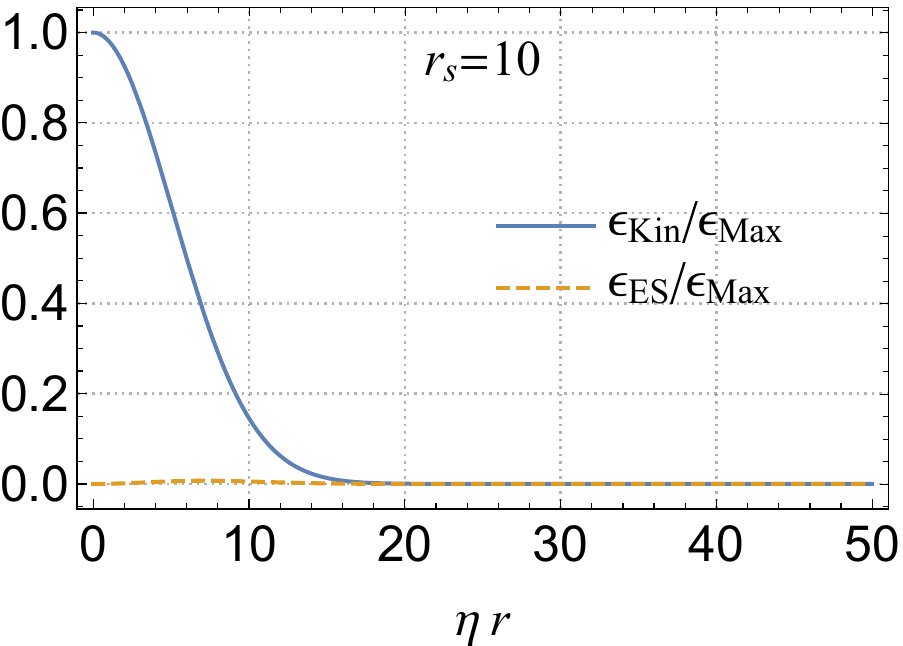}~~
\includegraphics[width=7cm]{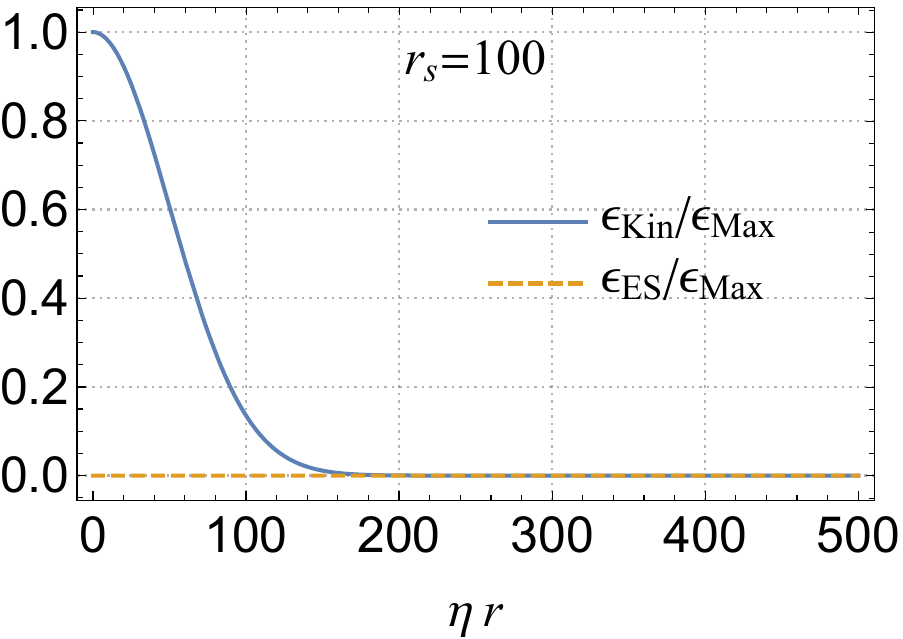}
\end{center}
\centering
\begin{minipage}[b]{0.9\hsize}
\caption{
The kinetic energy density, $\epsilon_\text{Kin}$,
and the electrostatic energy density, $\epsilon_\text{ES}$,
normalized by the maximum values of $\epsilon$
are plotted for $r_s= 0.1, 1, 10, 100$.
\label{fig:Gaussianenergydensity}
}
\end{minipage}
\end{figure}

%---------------------------------------
We inspect the energy density of the numerical solutions.
The components of energy density given by \eqref{eq:energy_density_1} and \eqref{eq:energy_density_2}
are shown in Fig.\ref{fig:Gaussianenergydensity}. The dominant components of energy density $\epsilon$
are $\epsilon_\text{Kin}$ and $\epsilon_\text{ES}$, while $\epsilon_\text{Elast}$ and $\epsilon_\text{Pot}$
are negligibly small in the present cases.

In the thin source case, $r_s\ll r_A$,
the electrostatic energy density dominates the total energy density
(see  $r_s=0.1$ case in the first panel of Fig.\ref{fig:Gaussianenergydensity} for example), i.e.,
\begin{align}
	\epsilon(r) \simeq \epsilon_\text{ES}(r)
		=\frac{1}{2}\biggl(\frac{d\alpha}{dr}\biggr)^2.
\label{eq:rs0p1originenergydensity}
\end{align}
In the near region $0\leq r\leq r_s$,
as shown in Appendix \ref{app:approx_sol_Gaussian},
the asymptotic behavior of the function $\alpha(r)$ near the origin is given by
\eqref{eq:Gaussian_thin_alpha}, i.e.,
\begin{align}
	\alpha(r) \sim \alpha_0 -\frac{\rho_0 r_s^2}{6} \left(\frac{r}{r_s}\right)^2,
\label{eq:originqsolution}
\end{align}
where $\alpha_0$ is the central value of $\alpha$.
Substituting \eqref{eq:originqsolution} into \eqref{eq:rs0p1originenergydensity},
the energy within $r_s$ is given by
\begin{align}
	E|_{r\leq r_s}
		\simeq \frac{4\pi\rho_0^2}{9} \int_0^{r_s}r^4dr
		=\frac{4 q^2}{45\pi^2r_s}.
\label{eq:rs0p1energy2}
\end{align}
In the region $r>r_s$, since $\alpha$ is given by \eqref{eq:q asymptotic inf2},
then the energy of the system \eqref{eq:energy2} in this range
can be written by
\begin{align}
	E|_{r>r_s}&\simeq 2\pi \int_{r_s}^{\infty}r^2\left(\frac{1}{r^2}
	+\frac{1}{rr_A}\right)^2\exp\left(-\frac{2r}{r_A}\right) dr
\cr
	&= 2\pi \exp\left(-\frac{2r_s}{r_A}\right)
		\left(\frac{1}{2r_A}+\frac{1}{r_s}\right) \simeq \frac{2\pi}{r_s}.
\label{eq:rs0p1energy}
\end{align}
Therefore, the total energy $E=E|_{r\leq r_s}+E|_{r>r_s}$ is proportional to $r_s^{-1}$.

%%%%%%%%%%%%%%%%%%%%%%%%%%%%%%%%%%%%%%%%%%%%%%%%%%%%%%%%%%%%%%%%%%%%%
%%%%%%%%%%%%%%%%%%%%%%%%%%%%%%%%%%%%%%%%%%%%%%%%%%%%%%%%%%%%%%%%%%%%%

In contrast, in a thick source case, $r_s\gg r_A$,
as shown in \eqref{eq:Gaussian_thick} of Appendix \ref{app:approx_sol_Gaussian},
we see that $f(r)\simeq \eta$ and $\alpha(r)=\rho_\text{ext}(r)/m_A^2$,
then the energy density becomes
\begin{align}
  \epsilon(r)&\simeq \epsilon_\text{Kin}(r) =e^2f^2\alpha^2
 =\frac1{2 m_A^2}\rho_0^2 \exp \left[-2\left(\frac{r}{r_s}\right)^2\right].
  \label{eq:xi100 energy density}
\end{align}
Therefore, the energy $E$ given by
\begin{align}
  E&=4\pi \int_0^{\infty}r^2 \epsilon(r)dr
    =\frac{q^2}{4\sqrt{2}\pi^{3/2} m_A^2 ~r_s^3}.
  \label{eq:xi100 total energy}
\end{align}
is proportional to $r_s^{-3}$.

%%%%%%%%%%%%%%%%%%%%%%%%%%%%%%%%%%%%%%%%%%%%%%%%%%%%%%%%%%%%%%%%
 \begin{figure}[H]
 \includegraphics[width=7cm]{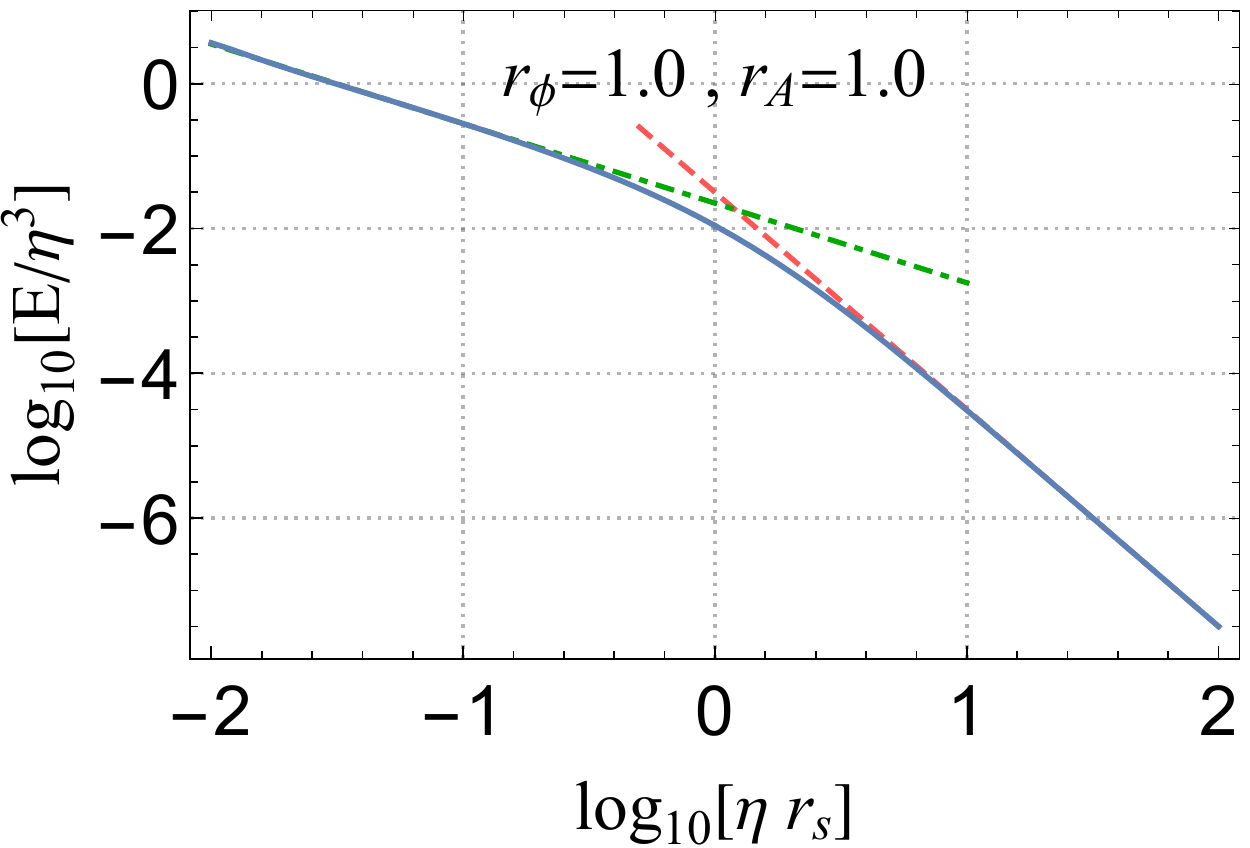}
 \includegraphics[width=7cm]{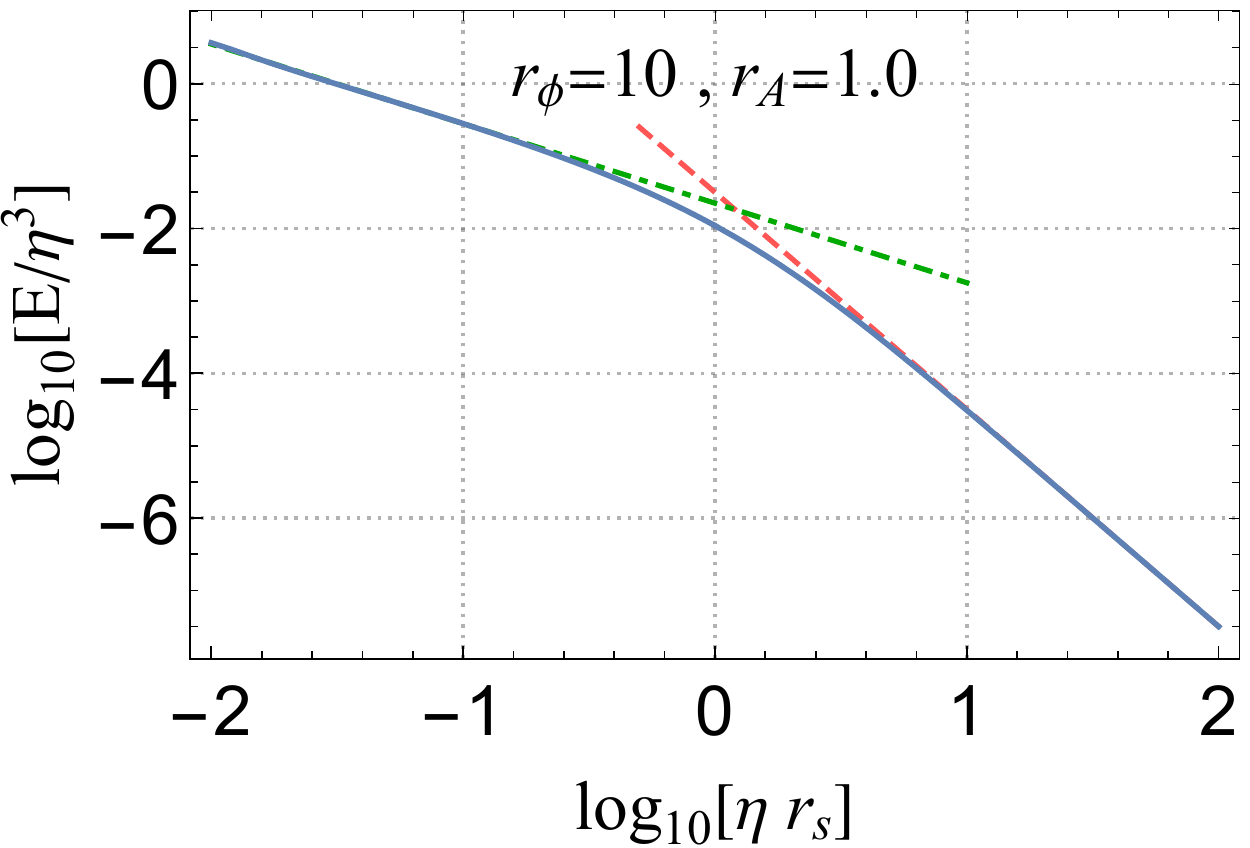}
\\
\\
 \includegraphics[width=7cm]{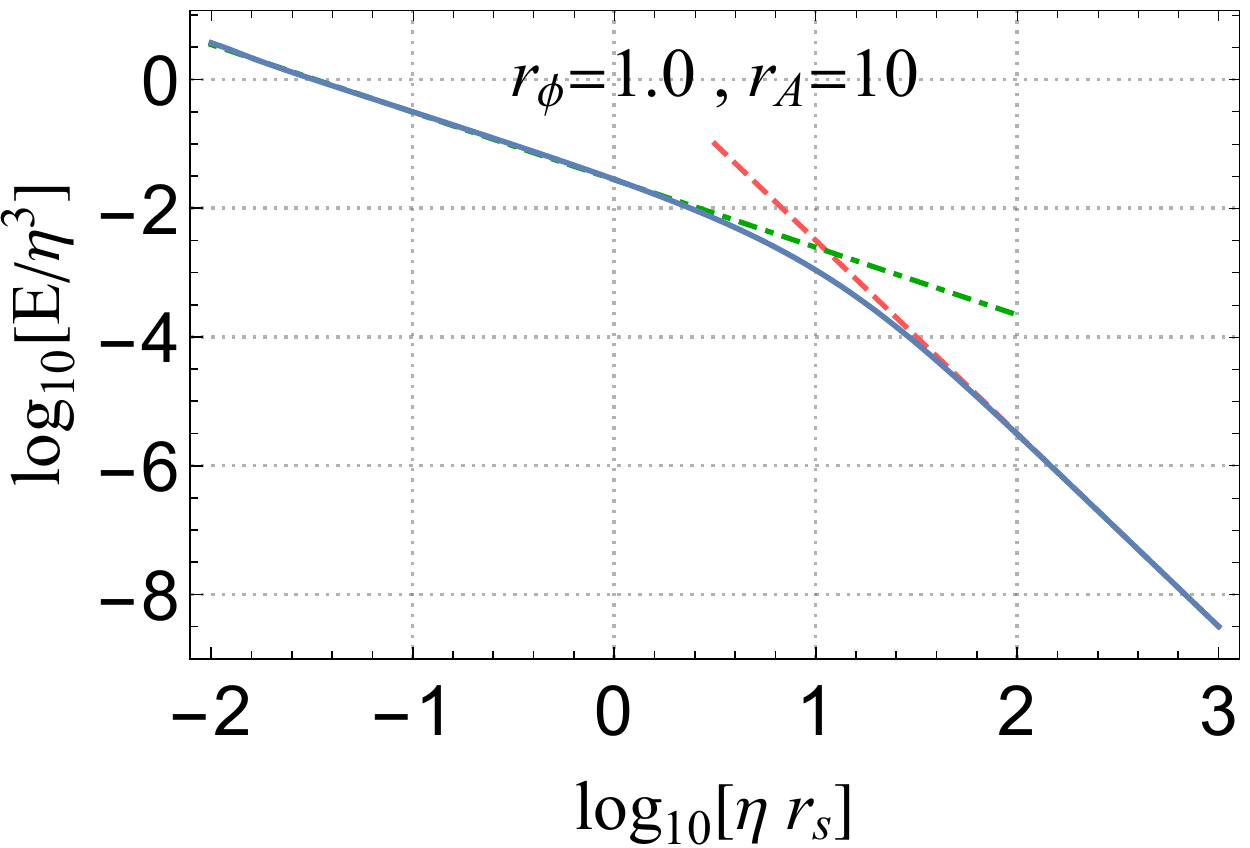}
 \includegraphics[width=7cm]{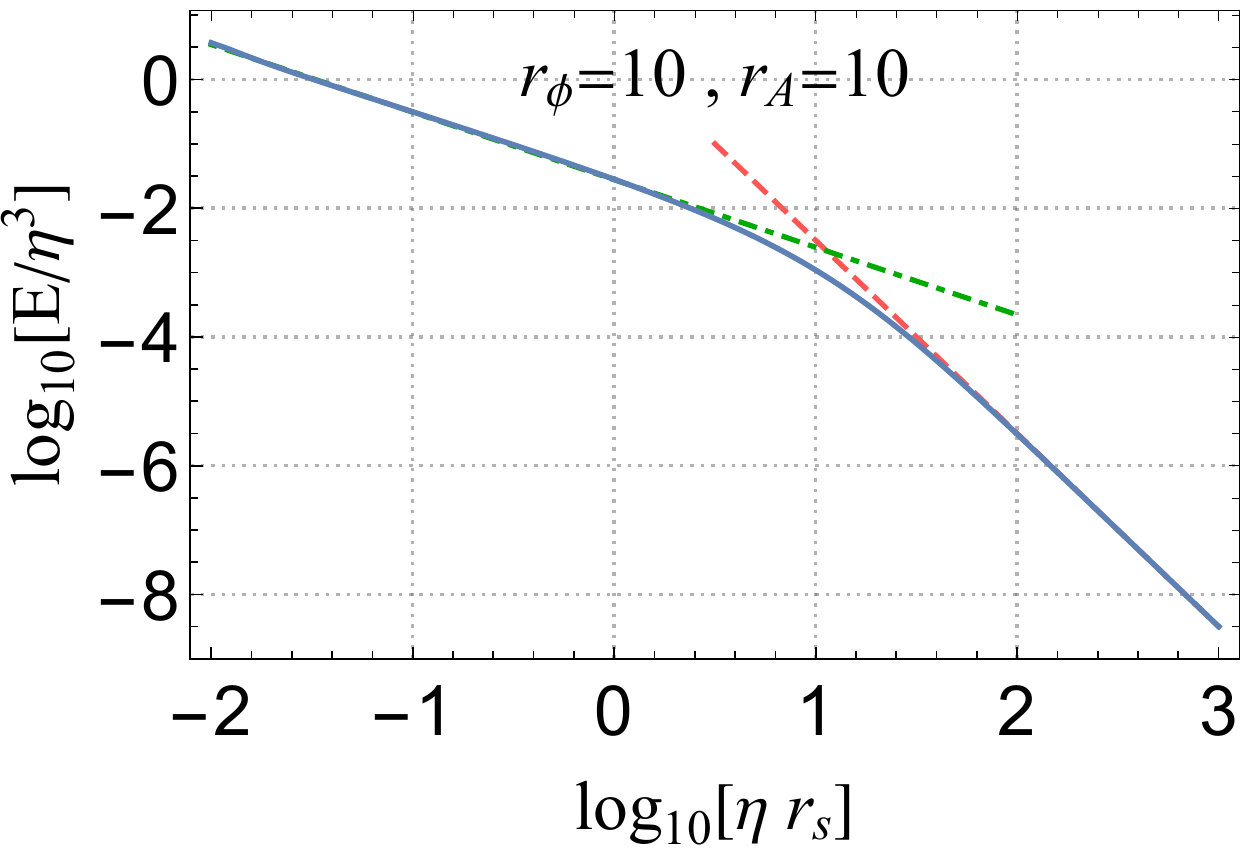}
\\
 \centering
 \begin{minipage}{0.9\hsize}
 \caption{
Log-log plot of the total energy $E$ versus width of the external charge $r_s$.
The broken line means $r_s^{-3}$ while the dot-dashed line means $r_s^{-1}$.
 }
 \label{fig:loglogenergy}
 \end{minipage}
 \end{figure}
%%%%%%%%%%%%%%%%%%%%%%%%%%%%%%%%%%%%%%%%%%%%%%%%%%%%%%%%%%%%%%%%

By numerical calculations for some values of the parameter sets $(r_{\phi}$, $r_A)$,
the energy $E$ is plotted as
a function of $r_s$ in Fig.\ref{fig:loglogenergy}.
In all cases, we see that $E \propto r_s^{-1}$ for small $r_s$, and
$E \propto r_s^{-3}$ for large $r_s$. The power index changes around $r_s=r_A$.

%----------------------------------------------
\subsection{Homogenious ball source}
%----------------------------------------------
As the second example of smoothly extended source,
we consider the ball of constant charge density expressed
by \eqref{eq:Homogenious source}.
As same as the Gaussian distribution case discussed above,
we set $e=1/\sqrt{2}$, and $\lambda=1$.
We fix the central charge density $\rho_0$,
and find numerical solutions for several values of $r_s$, radius of the ball,
and $\zeta_s$, surface thickness parameter.
Note that the total external charge is in proportion to $r_s^3$.

\begin{figure}[H]
\begin{minipage}{0.45\hsize}
\centering
\includegraphics[width=7cm]{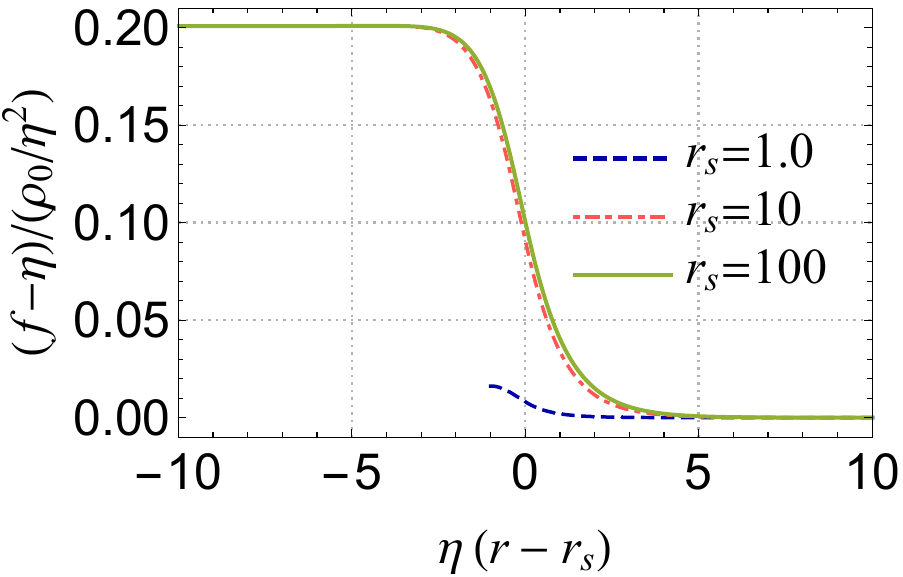}
\end{minipage}
\begin{minipage}{0.45\hsize}
\centering
\includegraphics[width=7cm]{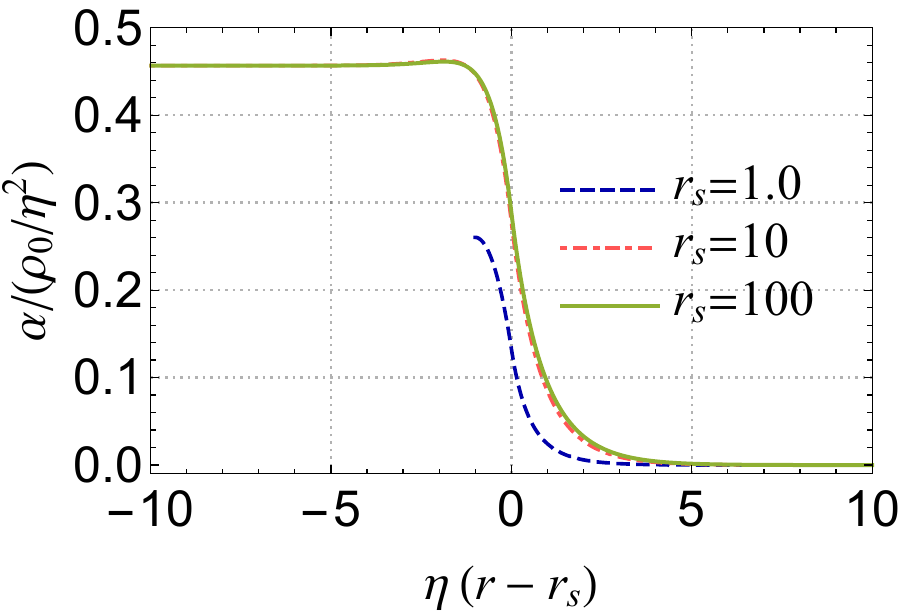}
\end{minipage}
\centering
\begin{minipage}{0.9\hsize}
\caption{
Behaviors of $f$ (left panel) and $\alpha$ (right panel)
in the vicinity of the ball surface $r_s$.
Three cases $r_s= 1, 10, 100$ are superposed.
}
\label{fig:tanhsolutionrsvicinity}
\end{minipage}
\end{figure}

By numerical calculations, $f(r)$ and $\alpha(r)$ with  $\rho_\text{ext}(r)$
are shown in the cases of $r_s=1, 10$, and $100$ for fixed surface thickness as $\zeta_s=0.01$
in Fig.\ref{fig:tanhsolution}.
As is shown in Appendix\ref{app:homogeneous_ball},
in the region $r <r_s-r_A$, where $\rho_\text{ext}\simeq \rho_0=const.$,
we see
\begin{align}
	f \simeq f_0
\quad\mbox{and}\quad
	\alpha \simeq \alpha_0,
\end{align}
where $f_0$ and $\alpha_0$ are constants given in Appendix\ref{app:homogeneous_ball}.
In the region $r \geq r_s+r_A$, where $\rho_\text{ext}\simeq 0$,
we see simply $f \simeq  \eta$ and	$\alpha \simeq 0$.
The functions $f$ and $\alpha$ change the values quickly in the vicinity of the ball surface
$r_s-r_A\leq r \leq r_s+r_A$.
The profiles of $f$ and $\alpha$ near the ball surface $r_s$ are almost identical if $r_s \gg r_A$
(see Fig.\ref{fig:tanhsolutionrsvicinity} ).

%%%%%%%%%%%%%%%%%%%%%%%%%%%%%%%%%%%%%%%%%%%%%%%%%%%%%%%%%%
\begin{figure}[H]
\includegraphics[width=5.5cm]{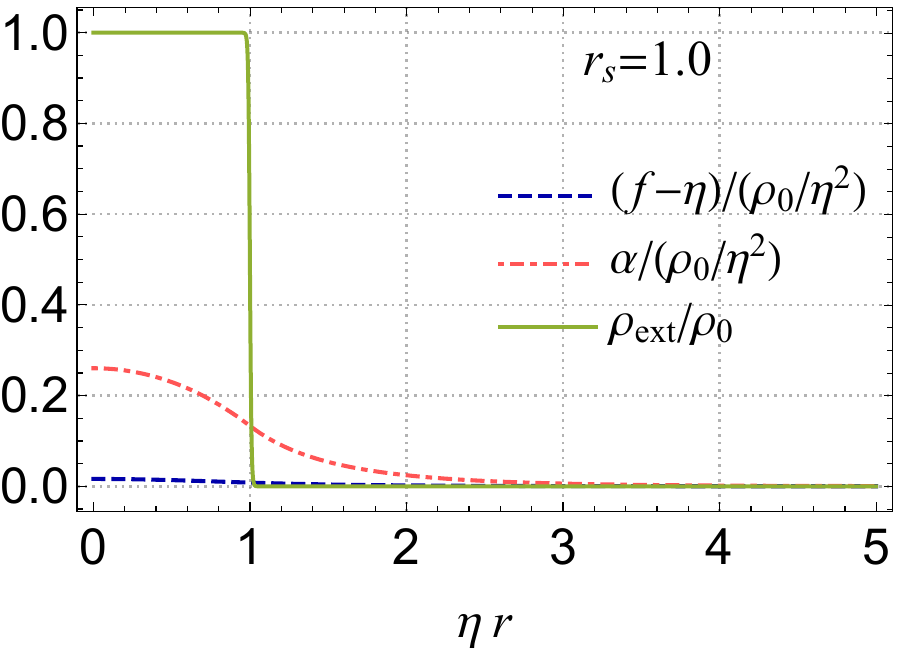}~~
\includegraphics[width=5.5cm]{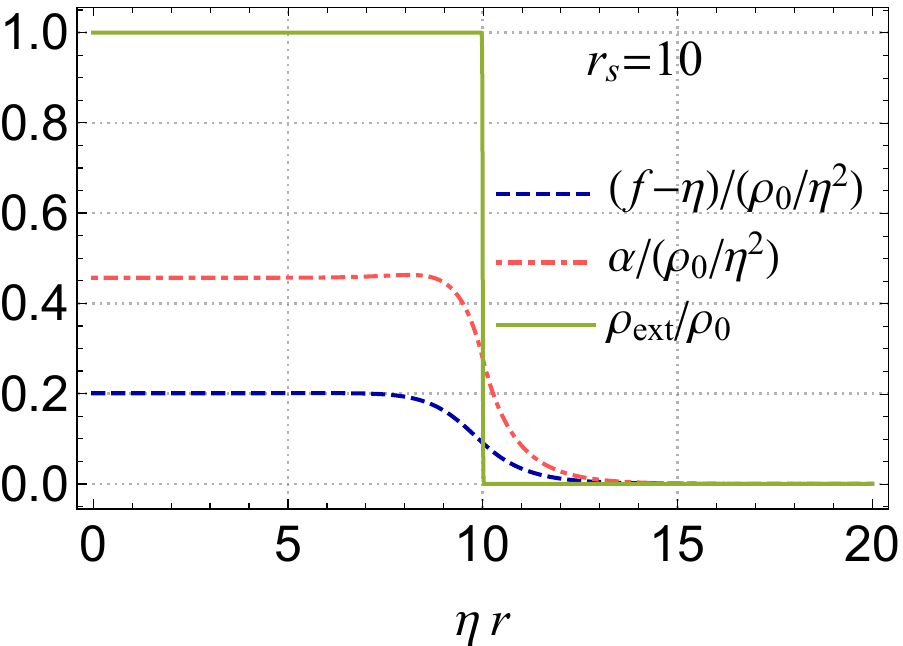}~~
\includegraphics[width=5.5cm]{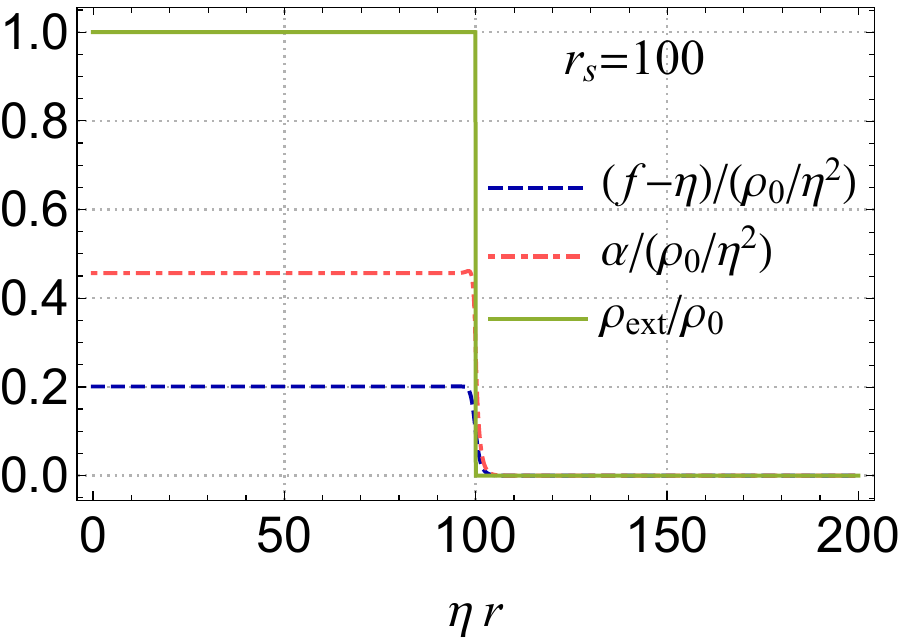}
\centering
\begin{minipage}{0.9\hsize}
\caption{
Behaviors of $f$, $\alpha$ and $\rho_\text{ext}$ for $r_s=1, 10, 100$
with fixed $\zeta_s=0.01$.
\label{fig:tanhsolution}
}
\end{minipage}
\end{figure}
%%%%%%%%%%%%%%%%%%%%%%%%%%%%%%%%%%%%%%%%%%%%%%%%%%%%%%%%%%%%%

%%%%%%%%%%%%%%%%%%%%%%%%%%%%%%%%%%%%%%%%%%%%%%%%%%%%%%%%%%%%%
%%%%%%%%%%%%%%%%%%%%%%%%%%%%%%%%%%%%%%%%%%%%%%%%%%%%%%%%%%%%%

Next, we consider variation of surface thickness $\zeta_s$ for fixed ball radius $r_s$.
The profile of the functions $f$ and $\alpha$, the charge density, and energy density
are shown in the cases of $\zeta_s=0.1, 1$, and $10$ for fixed ball radius as $r_s=100$
in Fig.\ref{fig:tanhsolution}.
Inside the homogeneous ball source, the induced charge density
cancels the external charge density except the vicinity of the ball surface.
In the thin ball surface case, $\zeta_s \ll r_A$, at the surface,
where $\alpha$ changes its value quickly, the induced charge exceeds the external
charge inside the surface, and vice versa outside.
Therefore, an electric double layer emerges at the surface of the ball.
For the thick surface case, $\zeta_s \gg r_A$, charge cancellation occur everywhere even at
the surface. Namely, the perfect screening occurs in this case.

\begin{figure}[H]
\includegraphics[width=5.5cm]{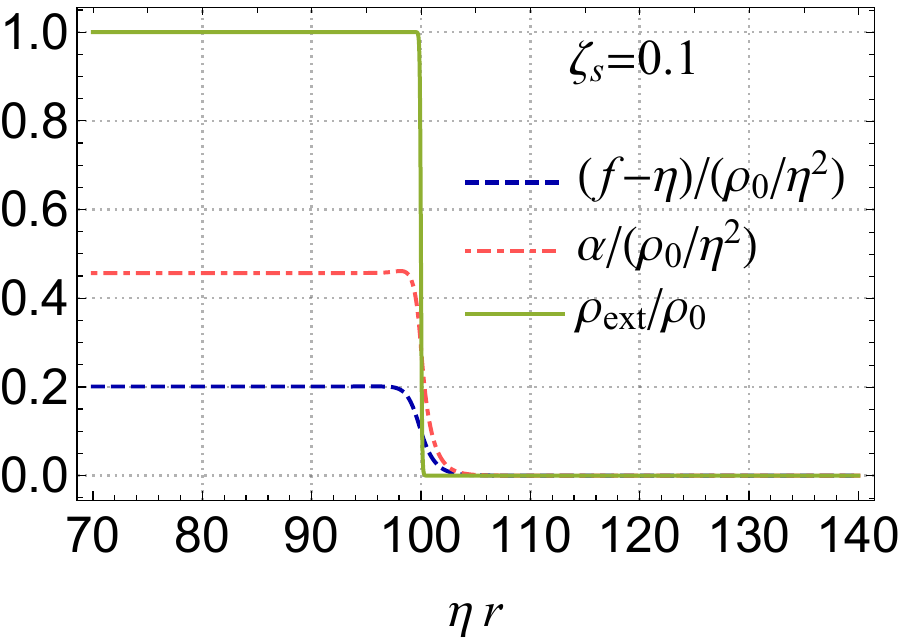}~
\includegraphics[width=5.5cm]{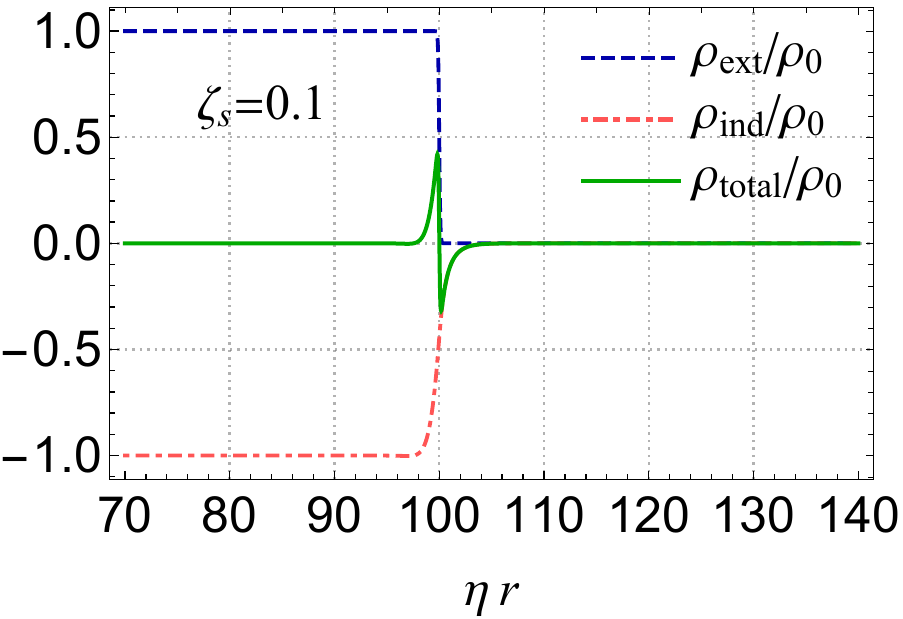}~
\includegraphics[width=5.5cm]{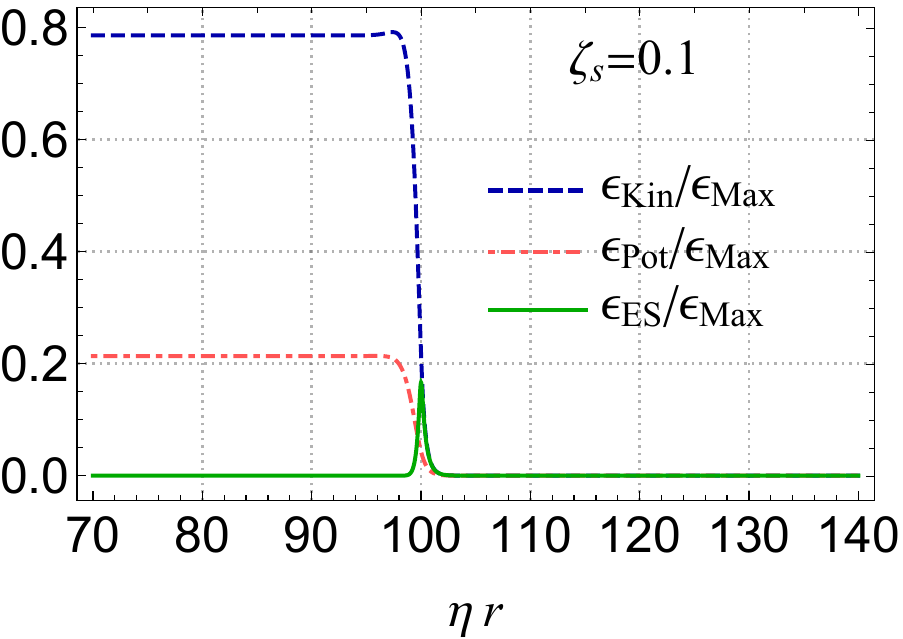}
\begin{minipage}{0.45\hsize}
\vspace{1cm}
\end{minipage}
\\
\includegraphics[width=5.5cm]{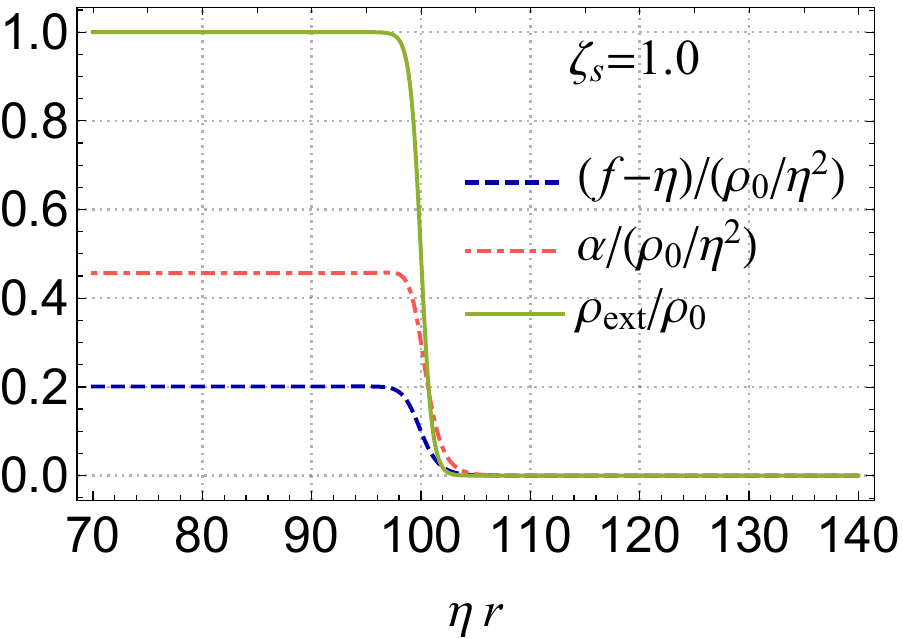}~
\includegraphics[width=5.5cm]{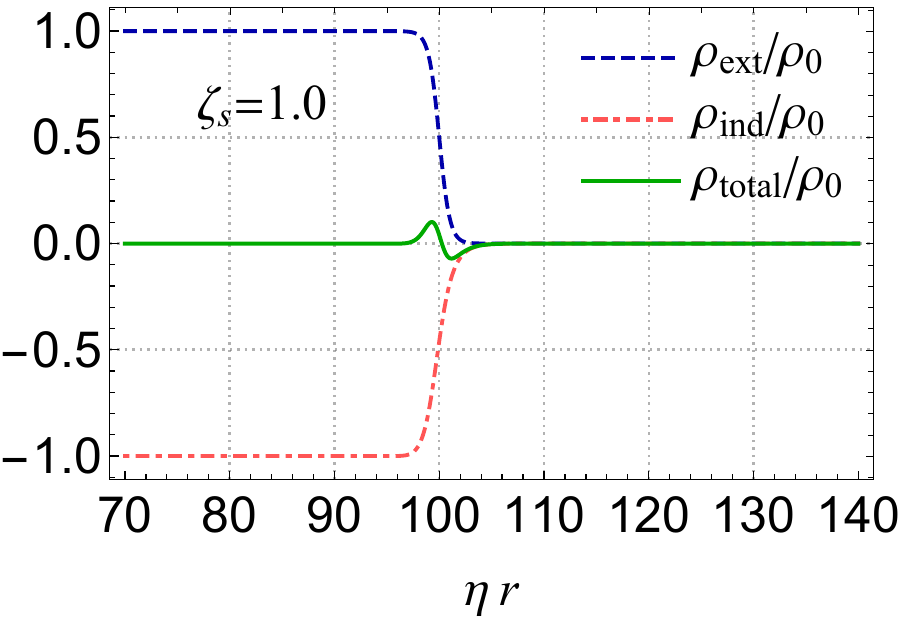}~
\includegraphics[width=5.5cm]{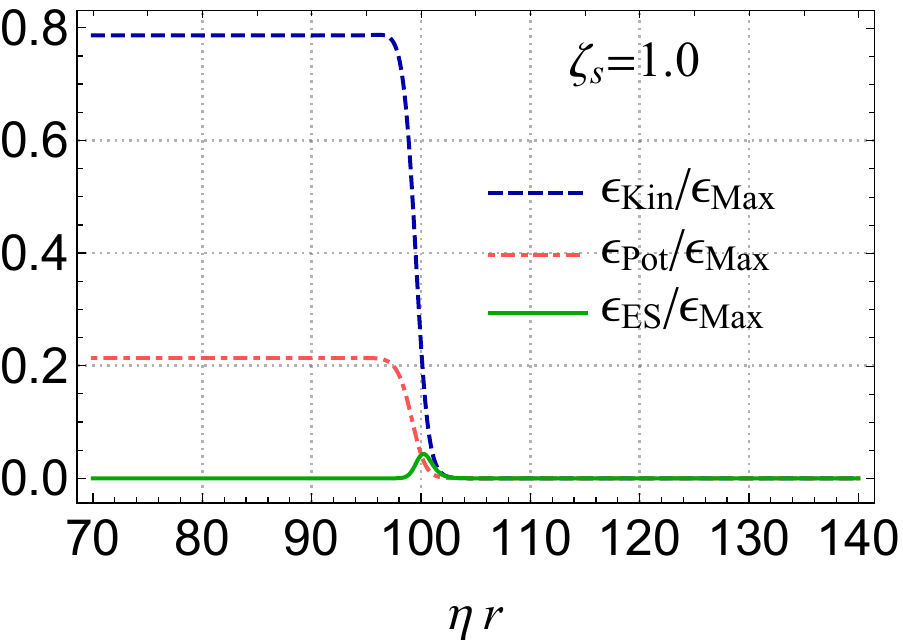}
\\ \\
\includegraphics[width=5.5cm]{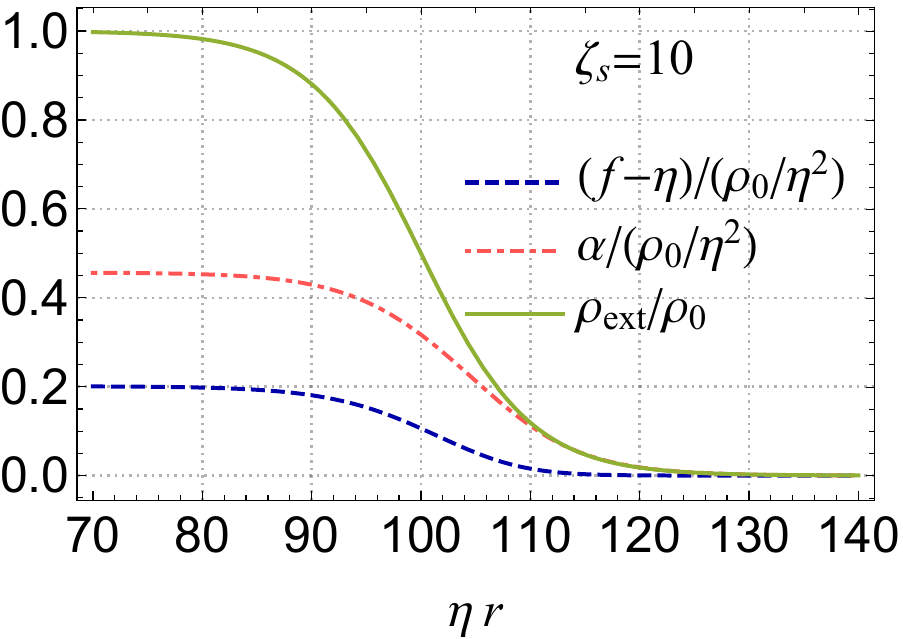}~
\includegraphics[width=5.5cm]{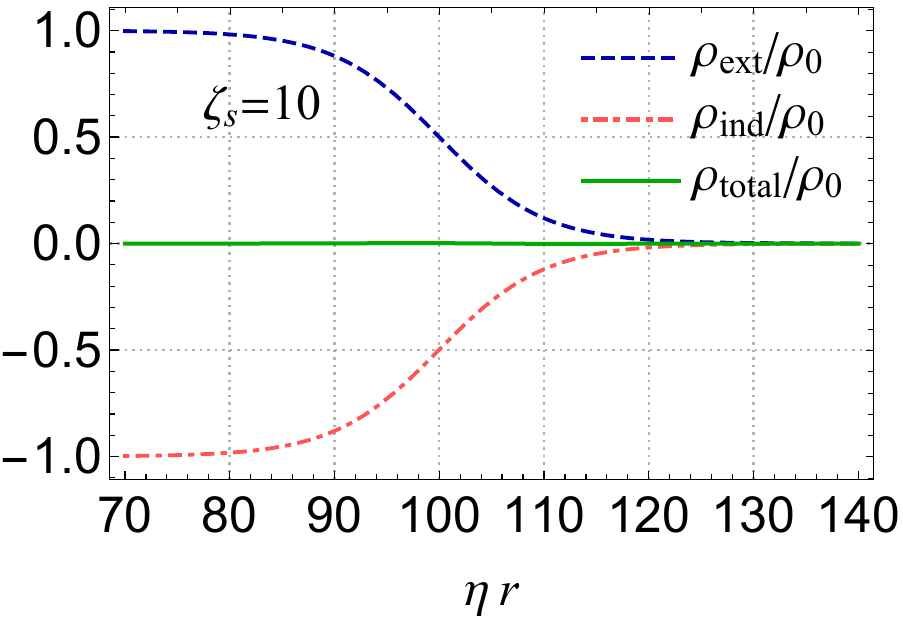}~
\includegraphics[width=5.5cm]{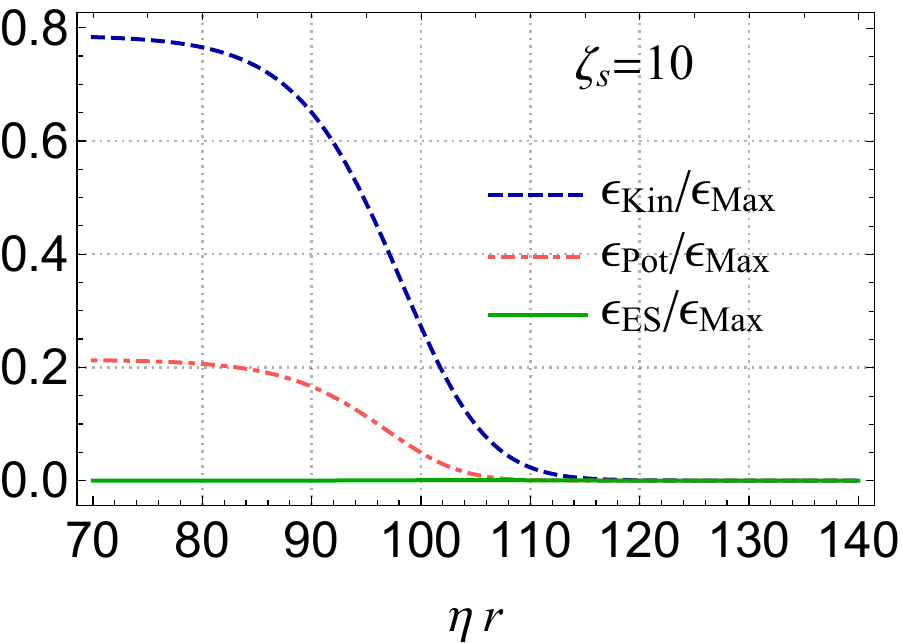}
\\
\caption{
Behaviors of $f$ and $\alpha$, the charge densities, the energy densities for various surface thickness
parameters $\zeta_s$.
The functions $f$ and $\alpha$ are shown together with $\rho_\text{ext}$ in the first column,
$\rho_\text{ext}$, $\rho_\text{ind}$ and $\rho_\text{total}$ are shown in the middle column,
and $\epsilon_\text{Kin}$, $\epsilon_\text{Pot}$ and $\epsilon_\text{ES}$ are shown in the right column.
The first row is for $\zeta_s=0.1$, the second for $\zeta_s=1$, and the last for $\zeta_s=10$.
\label{fig:variation_surface_thickness}
}
\end{figure}

The components of energy density given in \eqref{eq:energy_density_1} and \eqref{eq:energy_density_2}
are shown in Fig.\ref{fig:variation_surface_thickness}.
Inside the homogeneous ball, the kinetic energy dominate the energy density
and the electrostatic energy density caused by the electric double layer appears
at the neighborhood of surface for the thin surface case.

%%%%%%%%%%%%%%%%%%%%%%%%%%%%%%%%%
%%%%%%%%%%%%%%%%%%%%%%%%%%%%%%%%%%%%%%%%%%%%%%%%%%%%%%%%%%%%%%%%%%%%%%%%%%%%%%%

\newpage

%%%%%%%%%%%%%%%%%%%%%%%%%%%%%%%%%%%%%%
\section{Summary and Discussion}
%%%%%%%%%%%%%%%%%%%%%%%%%%%%%%%%%%%%%%

In this paper, we have studied the classical system that consists of a U(1) gauge field and a complex
Higgs scalar field with a potential that breaks the symmetry spontaneously.
We have presented numerical solutions in the presence of a smoothly extended external source
with a finite size.
Owing to the existence of the external source, counter charge cloud is induced
by the scalar and the vector fields.

We have investigated two extended external sources: Gaussian distribution sources and homogeneous
ball sources.
In the case of Gaussian distribution source, the profile of
the total charge within radius $r$, $Q(r)$, depends on the width of the external source, $r_s$.
In the thin source case, where $r_s$ is much smaller than the mass scale of the vector field,
$r_A =m_A^{-1}$,
non-vanishing peak of $Q(r)$ appears at a radius in the range $r<r_s$.
Then, the charge density is detectable in the region $r<r_s$.
The maximum value of $Q(r)$ is less than the total external charge, then the partial
screening occurs in a finite distance.
As $r$ increases, $Q(r)$ damps quickly, then the total charge screening occurs
by the induced charge cloud for a distant observer.
In the thick source case, where $r_s$ is much larger than $r_A$, $Q(r)$ is almost zero everywhere,
equivalently, $\rho_\text{total}$ almost vanishes everywhere.
In this case, the charge is perfectly screened so that the charge is not detectable anywhere.

In accordance with the induced charge cloud, the energy density of the fields is also induced around
the external source. In the thin source case, the electrostatic energy produced by the
non-vanishing total charge density appears dominantly.
In the thick source case, the kinetic energy, square of covariant time derivative
of the scalar field, dominates the energy density.
The total energy $E$ of the cloud depends on the thickness parameter $r_s$;
for the thin source, $E$ is proportional to $r_s^{-1}$,
while for the thick source, $E$ is proportional to $r_s^{-3}$.
The transition of the power index occur at $r_s \simeq r_A$.

For the homogeneous ball source, we have considered that the charge density is constant
within the ball radius, $r_s$, which is assumed to be much larger than $r_A$,
and the charge density varies with the surface thickness scale, $\zeta_s$,
at the ball surface.
We found that inside the ball, $r<r_s-r_A$, the amplitude of the scalar field
and the gauge field take constant values, respectively, and outside the ball, $r > r_s+r_A$,
the scalar field takes the vacuum expectation value and the gauge field vanishes.
At the ball surface, the both fields change their values quickly.
The external charge is canceled out by the induced charge cloud except the vicinity of ball surface.
In the thin surface case, $\zeta_s \ll r_A$, electric double layer appears at the ball surface.
In the thick surface case, $\zeta_s \gg r_A$, the charge cancellation occurs even at the ball surface,
namely, the perfect screening occurs.

The kinetic energy and the potential are main components of the energy density inside the ball.
For the thin surface case, the electrostatic component of the energy density by the electric double layer
appears at the ball surface.

In this paper, we have concentrated on the screening mechanism of external charge sources.
It is interesting that the external sources are replaced by charged non-topological solitons.
There exist non-topological soliton solutions of a complex scalar field where the conserved charges
are extended smoothly.
Can we expect the charge screening occurs on the non-topological solitons ?

In the studies of the non-topological solitons, typical profiles of charge density of solitons are
Gaussian distributions and homogeneous balls \cite{Friedberg_Lee_Sirlin,Coleman}.
Ungauged non-topological solitons are allowed
to have infinitely large mass \cite{Lee_Stein-Schabes_Watkins_Widrow},
while gauged solitons have upper limit of mass owing to repulsive
force between charges \cite{Lee_Stein-Schabes_Watkins_Widrow,Shi_Li}.
If a non-topological soliton exists
in a system consisting of a complex field, a Higgs scalar field, and a U(1) gauge field,
the charge screening of the soliton occurs as discussed in the present paper.
It would be expected that the charge screened soliton has a infinitely large mass.
If the solitons are spread wider than the mass scale of the gauge field, the perfect screening
would occur. This is a preferable property for dark matter in the universe.
We would report the existence of charge screened non-topological solitons, which would be an
interesting candidate for the dark matter, in the forthcoming paper \cite{Ishihara_Ogawa}.

\section*{Acknowledgements}
We would like to thank Dr. K.-i. Nakao, Dr. N. Maru, and Dr. R. Nishikawa for valuable discussion.
H.I. was supported by JSPS KAKENHI Grant Number 16K05358.

\newpage

%%%%%%%%%%%%%%%%%%%%%%%%%%%%%%%%%%%%%%%%%%%%%%%%%%%%%%%%%%%%%%%%
\appendix
\section{Asymptotic behaviors for the point source}
\label{app:A}
%%%%%%%%%%%%%%%%%%%%%%%%%%%%%%%%%%%%%%%%%%%%%%%%%%%%%%%%%%%%%%%%
We analyze asymptotic behaviors of the scalar and gauge fields governed
by \eqref{eq:eq_f} and \eqref{eq:eq_alpha} for a point source \eqref{eq:point_source}~\cite{Landau}.
The equations admit an exact solution
$\alpha(r)= q/(4 \pi r)$ and $f(r)=0$, the Coulomb solution.
However, this configuration does not minimize the energy \eqref{eq:energy2}, i.e.,
not the vacuum. To seek other solutions with non-vanishing $f(r)$,
we discuss asymptotic behavior of the fields near the point source and at infinity.

%---------------------------------------
\subsection{Near the point source}
%---------------------------------------

We assume that the asymptotic behavior of the fields in the
vicinity of the point source are given by
\begin{align}
	\alpha(r)\sim a_1 r^{\gamma},
\label{eq:q asymptotic}\\
	f(r) \sim b_1 r^{\beta}.
\label{eq:f asymptotic}
\end{align}
where $a_1$ and $b_1$ are non-vanishing constants.
Substituting these expression in  \eqref{eq:eq_f} and \eqref{eq:eq_alpha}, we obtain
\begin{align}
 	&\beta(\beta-1)r^{\beta-2}+2\beta r^{\beta-2}+e^2 a_1^2 r^{\beta+2\gamma}
		-\frac{\lambda}{2} b_1^2 r^{3\beta}+\frac{\lambda}{2}r^\beta=0,
 \label{eq:f asymptotic equation}
\\
 	&\gamma(\gamma-1)r^{\gamma-2}+2\gamma r^{\gamma-2}-2e^2 b_1 ^2r^{2\beta+\gamma}=0.
 \label{eq:q asymptotic equation}
\end{align}

First, we consider the case of $\beta>-1$.
In this case, we can ignore the third term in \eqref{eq:q asymptotic equation}, and obtain $\gamma=-1$.
By Gauss' integral theorem applied in a small volume including the point source,
we have
\begin{align}
	\alpha = \frac{a_1}{r}= \frac{q}{4\pi r}.
\end{align}
Since $\beta>-1$ and $\gamma=-1$,
the first three terms in \eqref{eq:f asymptotic equation} should compensate each other.
Then, we obtain
\begin{align}
 \beta =\frac{1}{2}\left(-1\pm \sqrt{1-4\kappa^2}\right)
\label{eq:f_power}
\end{align}
where $\kappa:=eq/4\pi$.

If $\kappa \leq 1/2$, $\beta$ is real number.
For the upper sign in \eqref{eq:f_power},
the elastic energy density $\epsilon_\text{Ela}$
defined in \eqref{eq:energy_density_1} is finite in the limit $r \to 0$, however
it diverges for the lower sign. Then, we take the positive sign in \eqref{eq:f_power}
for the power index of $f$.

If $\kappa>1/2$, $\beta$ becomes complex numbers
\begin{align}
 \beta=\frac{1}{2}\left(-1\pm i \sqrt{4\kappa^2-1}\right),
 \label{eq:f degree2}
\end{align}
then we have the real function $f(r)$ in the form
\begin{align}
	f(r)
%&=f_0^+r^{\beta_+}+f_0^-r^{\beta_-}\notag \\
%&=\frac{1}{\sqrt{r}}\left\{f_0^+\exp(i\sigma \log r)+f_0^-\exp(-i\sigma \log r)\right\}\notag \\
	&=\frac{b_1}{\sqrt{r}}\cos \left(\sigma \log r+c_1\right),
\label{eq:f asymptotic2}\\
\sigma:&=\frac{1}{2}\sqrt{4\kappa^2-1},
\end{align}
where $b_1$ and $c_1$ are constants.

In the case of $\beta \leq -1$, after some consideration, we see $b_1$ should vanish. Then, it is not
the case in which the expected solution exists.

%------------------------------
\subsection{Distant region}
%------------------------------
At spatial infinity, $\alpha$ approaches to zero, and $f$ does to $\eta$ asymptotically.
Then, in the distant region, we rewrite $f(r)$ as
\begin{align}
  	f(r)\sim \eta +\delta f(r),
 \label{eq:delta_f}
\end{align}
where $\delta f \to 0$ as $r \to \infty$.
Substituting \eqref{eq:delta_f} to \eqref{eq:eq_f} and \eqref{eq:eq_alpha},
we obtain a set of linear differential equations
\begin{align}
  	&\frac{d^2}{dr^2}\delta f+\frac{2}{r}\frac{d}{dr}\delta f-\frac{1}{r_\phi^2} \delta f=0,
 \label{f asymptotic eq inf}\\
 	&\frac{d^2}{dr^2}\alpha+\frac{2}{r}\frac{d}{dr}\alpha-\frac{1}{r_A^2} \alpha=0,
\label{q asymptotic eq inf}
\end{align}
where higher order terms in $\delta f$ and $\alpha$ are neglected.
Solving these equations, we obtain asymptotic behaviors of the functions as
\begin{align}
  	&\delta f(r)\sim \frac{b_2}{r}\exp \left(-\frac{r}{r_{\phi}}\right),
 \label{eq:f asymptotic inf3}\\
 	&\alpha(r)\sim \frac{a_2}{r}\exp \left(-\frac{r}{r_{A}}\right),
\label{eq:q asymptotic inf3}
\end{align}
where $b_2$ and $a_2$ are constants. These behaviors at the large distance are general if
the external source has a compact support around the origin.

\newpage

%%%%%%%%%%%%%%%%%%%%%%%%%%%%%%%%%%%%%%%%%%%%%%%%%%%%%%%%%%%%%%%%%%%%%%%%%%%%%%%%%%%%%%
\section{Approximate solutions for the Gaussian distribution sources}
\label{app:approx_sol_Gaussian}
%%%%%%%%%%%%%%%%%%%%%%%%%%%%%%%%%%%%%%%%%%%%%%%%%%%%%%%%%%%%%%%%%%%%%%%%%%%%%%%%%%%%%%
First, we consider the thin source case, $r_s\ll r_A$.
As shown in the first panel of Fig.\ref{fig:Gaussian_configuration} and Fig.\ref{fig:Gaussianchargedensity}
for the case $r_s=0.1$ as an example,
we see
\begin{align}
	|\rho_\text{ind}|=2e^2 f^2 \alpha\ll \rho_\text{ext}
		\quad \text{and}\quad
	\eta^2< f^2 \ll \alpha^2
\end{align}
in the near region, $0\leq r\leq r_s$.
Then,  \eqref{eq:eq_f} and \eqref{eq:eq_alpha} reduces to
\begin{align}
	&\frac{d^2 f}{dr^2}+\frac{2}{r}\frac{df}{dr} +\alpha^{2} f =0,
\\
	&\frac{d^2\alpha}{dr^2}+\frac{2}{r}\frac{d\alpha}{dr}
		+\rho_0 \exp\left[-\left(\frac{r}{r_s}\right)^2\right]=0.
%	+\frac{q}{(\sqrt{\pi}r_s)^{3}}\exp\left\{-\left(\frac{r}{r_s}\right)^2\right\}=0,
\end{align}
in this region. We easily find a set of approximate solutions that
satisfies the boundary condition \eqref{eq:BC_origin} in the expansion form
\begin{align}
	&\alpha(r)= \alpha_0 -\frac{\rho_0 r_s^2}{6} \left(\frac{r}{r_s}\right)^2
		+ {\cal O}\left(\frac{r}{r_s}\right)^4,
\label{eq:Gaussian_thin_alpha}
\\
	&f(r)= f_0 \left(1 -\frac{\alpha_0^2 r_s^2}{6} \left(\frac{r}{r_s}\right)^2
		+ {\cal O}\left(\frac{r}{r_s}\right)^4 \right),
\label{eq:Gaussian_thin_f}
\end{align}
where $\alpha_0:=\alpha(0)$ and $f_0:=f(0)$.

In the far region, $r\gg r_s$, the functions $f$ and $\alpha$ take the same forms
of the point source case.
The constants $\alpha_0$ and $f_0$ should be
adjusted so that the solutions are smoothly connected from the near region to
the far region.

Next, we consider the thick source case, $r_s \gg r_A$.
Since the source is spread widely, the variation of the external charge density
is very small. Accordingly, the variation of the functions $f$ and $\alpha$ are
also small as is seen in the last panel
of Fig.\ref{fig:Gaussian_configuration} as an example.
Then the derivative terms in \eqref{eq:eq_f} and \eqref{eq:eq_alpha}
can be negligible, and we have
\begin{align}
	&-2 e^2 \alpha^2 +\lambda(f^2-\eta^2) =0 ,
%	&f^2-\eta^2 \simeq \frac{2 e^2 \alpha^2}{\lambda}.
\label{eq:thick_f}
\\
	&\rho_\text{ind} = -\rho_\text{ext}.
%	&2 e^2 f^2 \alpha = \rho_\text{ext}.
%	&\alpha \simeq \frac{\rho_\text{ext}}{2 e^2 f^2}.
\label{eq:thick_alpha}
\end{align}
If the external charge density $\rho_\text{ext}$ is small such that
\begin{align}
%	\rho_\text{ext} \ll m_A m_\phi \eta ,
	\rho_\text{ext} \ll  \frac{\eta}{r_A~ r_\phi} ,
\end{align}
we have
\begin{align}
	f \simeq \eta, \quad \text{and}\quad \alpha \simeq \frac{\rho_\text{ext}}{m_A^2}.
\label{eq:Gaussian_thick}
\end{align}
This behavior is seen in the last panel of Fig.\ref{fig:Gaussian_configuration}.

%%%%%%%%%%%%%%%%%%%%%%%%%%%%%%%%%%%%%%%%%%%%%%%%%%%%%%%%%
\section{Approximate solutions for the homogeneous ball sources}
\label{app:homogeneous_ball}
%%%%%%%%%%%%%%%%%%%%%%%%%%%%%%%%%%%%%%%%%%%%%%%%%%%%%%%%

In the homogeneous ball sources with $r_s\gg r_A$, except the vicinity of
the ball surface, $r_s-r_A < r < r_s+r_A$,
$f$ and $\alpha$ are almost constants. We can approach approximately to this simple behaviors.

In the region $r <r_s-r_A$, where $\rho_\text{ext}\simeq \rho_0=const.$,
since the derivative terms in \eqref{eq:eq_alpha} and \eqref{eq:eq_f} can be ignored
for the solutions that satisfy the boundary condition \eqref{eq:BC_origin},
then $f$ and $\alpha$ take constant values.
The equations of motion reduce to
\begin{align}
	&e^2f\alpha^2-\frac{\lambda}{2}f(f^2-\eta^2)=0,
\\
	&-2e^2f^2\alpha+\rho_0=0.
\end{align}
By solving these coupled algebraic equations, we obtain
\begin{align}
	f^2 &\simeq f_0^2 = \frac13 \eta^2 \left[
		1+ \left( 1+X+\sqrt{X(2+X)} \right)^{1/3}
				+ \left( 1+X+\sqrt{X(2+X)} \right)^{-1/3} \right],
\\
	\alpha &\simeq \alpha_0 =\frac{\rho_0}{2 e^2 f_0^2},
\end{align}
where $X$ is the constant defined by
\begin{align}
	X:= \frac{27 r_A r_\phi}{2 \eta}\rho_0 .
\end{align}
In the region $r \geq r_s+r_A$, where $\rho_\text{ext}\simeq 0$,
we have simply $f \simeq  \eta$ and	$\alpha \simeq 0$. The fields $f$ and $\alpha$
change their values quickly in the vicinity of the ball surface.

If $\rho_0 \ll \eta/(r_\phi r_A )$,
a global solution can be obtained approximately.
In this case, as same as the Gaussian source case, $f(r)\sim \eta$.
Moreover, if $\zeta_s \ll r_A$, the equation of the gauge field can be reduced to
\begin{align}
   \frac{d^2\alpha}{dr^2}+\frac{2}{r}\frac{d\alpha}{dr}-m_A^2\alpha+\rho_0\theta(r_s-r)=0,
\label{eq:q asymptotic eq tanh}
\end{align}
where $m_A=r_A^{-1}$. This is the Proca equation for a homogenious ball source.

In the region $r < r_s$, since $\rho_\text{ext}(r)=\rho_0$,
we have a solution
\begin{align}
  \alpha(r)
  =\frac{C_1}{r}\sinh \left(r/r_A \right) + \rho_0 r_A^2,
\label{eq:alpha_inside}
\end{align}
while in the region $r > r_s$,
we have
\begin{align}
 	\alpha(r) =\frac{C_2}{r}\exp \left(- r/r_A \right),
\label{eq:alpha_outside}
\end{align}
where $C_1$ and $C_2$ are constants that should be determined by continuity.
This is achieved by junction conditions for \eqref{eq:alpha_inside} and \eqref{eq:alpha_outside}
at the surface $r=r_s$ as
\begin{align}
  	C_2\exp(-r_s/r_A )&=C_1\sinh(r_s/r_A )+\rho_0 r_s r_A^2,
\label{eq:junction condition1}
\\
	-C_2\exp(-r_s/r_A )\left(1+\frac{r_s}{r_A}\right)
		&=C_1\left[\frac{r_s}{r_A}\cosh( r_s/r_A) -\sinh(r_s/r_A) \right].
\label{eq:junction condition2}
\end{align}
By solving \eqref{eq:junction condition1} and \eqref{eq:junction condition2}, we obtain
\begin{align}
	C_1&=-\frac{\rho_0 r_A^3 \left(1+\frac{r_s}{r_A}\right)}{\cosh(r_s/r_A)+\sinh(r_s/r_A)},
\label{eq:junction condition3}
\\
	C_2&=\frac{\rho_0 r_A^3 \exp(r_s/r_A ) \left[\frac{r_s}{r_A} \cosh(r_s/r_A)-\sinh(r_s/r_A)\right]}
	{\cosh(r_s/r_A)+\sinh(r_s/r_A)}.
\label{eq:junction condition4}
\end{align}

Another simple case is that of $\zeta_s \gg r_A$.
The derivative terms in \eqref{eq:eq_alpha} and \eqref{eq:eq_f} can be ignored everywhere
for the solutions that satisfy the boundary conditions \eqref{eq:BC_origin} and \eqref{eq:BC_infty},
then the equations of motion reduce to
\begin{align}
	&e^2f\alpha^2-\frac{\lambda}{2}f(f^2-\eta^2)=0,
\\
	&-2e^2f^2\alpha+\rho_\text{ext}=0.
\end{align}
Therefore,
\begin{align}
	f^2(r) &\simeq \frac13 \eta^2 \left[
		1+ \left( 1+Y(r)+\sqrt{Y(r)(2+Y(r))} \right)^{1/3}
				+ \left( 1+Y(r)+\sqrt{Y(r)(2+Y(r))} \right)^{-1/3} \right],
\\
	\alpha(r) &\simeq \frac{\rho_\text{ext}(r)}{2 e^2 f(r)^2},
\end{align}
where $Y(r)$ is the function defined by
\begin{align}
	Y(r):= \frac{27 r_A r_\phi }{2 \eta}\rho_\text{ext}(r) .
\end{align}

\newpage

%%%%%%%%%%%%%%%%%%%%%%%%%%%%%%%%%%%%%%%%%%%%%%%%%%%%%%%%%%%%%%%%%%%%%%%%%%%

\end{document}